\documentclass[12pt]{article}
\usepackage[a4paper, left=1in, right=1in, top=1in, bottom=1in]{geometry}
\usepackage{amsmath}
\usepackage{mathrsfs}
\usepackage{dutchcal}
\usepackage{amssymb}
\usepackage{amsfonts}
\usepackage{graphicx}
\usepackage{subcaption}
\usepackage{hyperref}
\usepackage{geometry}
\usepackage{float}
\usepackage{tikz}
\usepackage{tikz-3dplot} 
\usepackage{authblk}
\usepackage{ulem}
\usepackage[numbers,sort&compress]{natbib}
\usepackage{xcolor}
\usepackage{rotating}

\usepackage{mathrsfs}

\graphicspath{{Figures/}}

\usepackage[compat=1.1.0]{tikz-feynman}
\tikzfeynmanset{every diagram={small}}

\newcommand{\mA}{\mathcal{A}}
\newcommand{\mM}{\mathscr{M}}

\begin{document}
\title{\bf{Study of $CP$ violation in $\Lambda_b^0\rightarrow N^*M$ decays with the final-state rescattering mechanism}}

\author{Hui-Qiang Shang$^a
$, 
Tian-Liang Feng$^b
$, 
Jing Gao$^c
\footnote{Contact author: gao@hiskp.uni-bonn.de}
$, 
Qin Qin$^a
$
and Fu-Sheng Yu$^b
$}

\affil{\small $^a$\, School of Physics, Huazhong University of Science and Technology, Wuhan 430074, China}
\affil{\small $^b$\, Frontiers Science Center for Rare Isotopes and School of Nuclear Science and Technology, Lanzhou University, Lanzhou 730000, China}
\affil{\small $^c$\, Helmholtz-Institut f\"{u}r Strahlen- und Kernphysik and Bethe Center for Theoretical Physics, Universit\"{a}t Bonn, D-53115 Bonn, Germany}
\date{}

\maketitle

\begin{abstract} 
In this work, we investigate the charmless nonleptonic two-body  $\Lambda_b$ decays within the framework of final-state rescattering mechanism. Using this framework, we evaluate the relevant hadronic triangle diagrams by performing the loop integrations, retaining both their absorptive and dispersive parts.
Based on the established formalism, we analyze the $\Lambda_b \to N^*(1535,1520)M$ decay processes with $M =K_S, K^*_0(700)$, $f_0(500,980), \rho(770), \bar{K}^{*0}$, $\phi$, and predict various physical observables, such as their branching ratios, direct and partial-wave $CP$ asymmetries, as well as decay asymmetry parameters. These two-body decay processes are expected to contribute primarily to the subsequent four-body decay channels, such as $\Lambda_b^0 \to p\,\pi^-\,\pi^+\,\pi^-$, whose $CP$ asymmetry measurements will be accessible at the LHCb experiment.
\end{abstract}

\section{INTRODUCTION}
$CP$ violation plays a crucial role in understanding the matter–antimatter asymmetry in the Universe~\cite{Sakharov:1967dj}, testing the validity of the Standard Model (SM), and probing possible new physics beyond the SM. 
In the SM, $CP$ violation is explained through the well-known quark mixing  mechanism, originally proposed by the Kobayashi and Maskawa,  which is attributed to the complex phase in the Cabibbo-Kobayashi-Maskawa (CKM) matrix \cite{Cabibbo:1963yz, Kobayashi:1973fv}. However, compared to the observed baryon asymmetry in the universe~\cite{Planck:2015fie}, the $CP$ violation predicted by the SM is far too small. This phenomenon strongly indicates that additional sources of $CP$ violation beyond SM are required.

The CKM mechanism has been firmly validated in meson systems. The first experimental discovery of $CP$ violation was observed in $K$ meson decays \cite{Muller:1960ph, Christenson:1964fg, KTeV:1999kad}. Subsequently, $CP$-violating effects were identified in $B$ and $D$ meson decays \cite{BaBar:2001ags, Belle:2001zzw, BaBar:2004gyj, Belle:2004nch, LHCb:2013syl, LHCb:2019hro}, thus further corroborating the validity of the theory. Recently, the LHCb collaboration has reported  the first observation of the significant $CP$ violation in $\Lambda_b^0 \rightarrow p\pi^+K^-\pi^-$ decay channel \cite{LHCb:2025ray}, and the measured $CP$ asymmetry is given by
\begin{equation}
    \begin{aligned}
        \mathcal{A}_{CP} = \frac{\Gamma - \bar{\Gamma}}{\Gamma + \bar{\Gamma}} = (2.45 \pm 0.46 \pm 0.10)\% \ ,
    \end{aligned}
\end{equation}
where $\Gamma$ and $\bar{\Gamma}$ denote the decay width of the $\Lambda_b^0 \rightarrow p\pi^+K^-\pi^-$ process and its $CP$-conjugate process, respectively. This is the first time that LHCb collaboration reported the definitive observation of $CP$ violation in baryon decays, confirming a long-sought phenomenon in particle physics. This breakthrough result represents a pivotal advancement in elucidating fundamental symmetry-breaking mechanisms and provides a new platform for testing the SM predictions and probing new physics beyond the SM.

Theoretical understanding and predictions of $CP$ violation in baryon decays are also of great importance. Notably, $CP$ asymmetries in $\Lambda_b\to pK^-\pi^+\pi^-$ have been measured in four distinct regions of phase space. In the channel $\Lambda_b\to R(p\pi^+\pi^-)K^-$, where the $p\pi^+\pi^-$ system has low invariant mass, a $CP$ asymmetry of $(5.4\pm0.9)\%$ has been observed with a significance of $6.0\,\sigma$ \cite{LHCb:2025ray}. This result aligns well with theoretical predictions based on a rescattering mechanism using $N\pi\to N\pi\pi$ scattering data, which yields $A_{CP}(\Lambda_b\to R(p\pi^+\pi^-)K^-) = (5.6\!-\!5.9)\%$ \cite{Wang:2024oyi}. This achievement demonstrates the promising potential of the final-state rescattering approach for predicting $CP$ asymmetries in $b$-baryon decays.  
In the same analysis  \cite{LHCb:2025ray}, the region corresponding to $\Lambda_b\to R(p\pi^-)R(K^-\pi^+)$ has been measured, yielding $A_{CP} = (2.7\pm0.8)\%$. In baryon decays, excited nucleon states typically contribute dominantly, with the lowest-lying excitations being $N^*(1520)$ and $N^*(1535)$. Thus, the main resonant contributions in this process are $N^*(1520)/N^*(1535)\to p\pi^-$ and $\bar{K}^{*0}/\bar{K}_0^*(700)\to K^-\pi^+$.  
At present, excited states remain challenging to study within the perturbative QCD factorization approach \cite{Han:2024kgz}, due to the lack of light-cone distribution amplitudes for excited nucleons. However, such states can be effectively investigated using final-state-interaction rescattering mechanisms.

In this work, we extend the final-state rescattering framework of Ref.~\cite{Duan:2024zjv} from channels containing ground-state baryons to 14 channels of the form $\Lambda_b^0\to N^*(1535,1520)M$, where the final state consists of an excited nucleon $N^*$ and a meson \(M\), with $M =K_S, K^*_0(700)$, $f_0(500,980)$, $\rho(770)$, $\bar{K}^{*0}$, and $\phi$. We adopt the loop-integration prescription and the two model parameters from Ref.~\cite{Duan:2024zjv}. The relevant hadronic triangle diagrams are evaluated with both their absorptive and dispersive parts retained.
The extension to the spin-$1/2$ resonance $N^*(1535)$ and the spin-$3/2$ resonance $N^*(1520)$ also requires the corresponding excited-nucleon strong-interaction vertices and spin-dependent amplitude structures. We calculate branching ratios, direct and partial-wave $CP$ asymmetries, and decay-asymmetry parameters for these channels. For the $N^*(1520)V$ modes, we also evaluate complementary $T$-odd and $T$-even angular $CP$ asymmetries. These observables provide information on the relative magnitudes and phases of the decay amplitudes and their partial-wave and helicity structures. After the subsequent resonance decays, the $N^*M$ modes  can constitute quasi-two-body components of experimentally accessible four-body $\Lambda_b$ decays. The present predictions can therefore provide input for amplitude analyses of resonance-dominated regions of these decays.

The structure of this paper is as follows. In Sec. II, we provide a concise overview of the theoretical framework employed in this work, including the naive factorization hypothesis, the quark-level topological diagram approach, and the final-state rescattering mechanism. The fundamental input parameters and numerical results are displayed in Sec. III. In Sec. IV, we give a brief summary of this work and an outlook for future studies. The effective Lagrangians, the corresponding strong interaction vertices, and relevant expressions of the triangle diagrams and amplitudes are collected in the Appendixes.

\section{THEORETICAL FRAMEWORK}
In this section, we introduce the theoretical framework that is used to calculate the amplitudes of charmless two-body nonleptonic $\Lambda^0_b$ decays. Based on the weak effective Hamiltonian, the topological diagrams at partonic level for each decay channel are obtained. The naive factorization  hypothesis is adopted to evaluate the short-distance hadronic matrix elements, and the rescattering mechanism is employed to address the long-distance final-state interactions.

\subsection{The effective Hamiltonian and short-distance contributions}
The charmless two-body nonleptonic $\Lambda^0_b$ decay processes involve at least three distinct scales: $m_W \gg m_b \gg \Lambda_{\text{QCD}}$. To systematically account for the physics at different scales, it is advantageous to employ the effective field theory. By integrating out the heavy degrees of freedom associated with the electroweak scale (e.g., the $W$ boson) and performing the operator product expansion, we obtain an effective weak Hamiltonian that governs $b$-quark decays at the scale $\mu= m_b$. Therefore, all the short-distance contributions above the scale $m_b$ are encapsulated in the Wilson coefficients, which are perturbatively calculable and independent of the specific hadronic initial and final states. Moreover, the nonperturbative long-distance effects are encoded in the hadronic matrix elements of local operators.
The effective weak Hamiltonian for the $b \to u$ transition at tree level and the $b \to q$ transition ($q = d, s$) at loop level takes the form~\cite{Buchalla:1995vs, Lu:2009cm}:
\begin{equation}
    \begin{aligned}
        \mathcal{H}_{eff} = \frac{G_F}{\sqrt{2}} \{V_{ub}V_{uq}^*[C_1(\mu)O_1(\mu) + C_2(\mu)O_2(\mu)] - V_{tb}V_{tq}^*[\sum_{i = 3}^{10}C_i(\mu)O_i(\mu)] \}+H.c. \ ,
    \end{aligned}
\end{equation}
where $G_F$ is the Fermi coupling constant, $V_{ij}$ are the elements of the CKM matrix, $C_i(\mu)$ are the Wilson coefficients evaluated at the renormalization scale $\mu$, and $O_i(\mu)$ denote the corresponding local four-quark operators. The explicit forms of $O_i(\mu)$ and the numerical values of $C_i(\mu)$ can be found in~\cite{Lu:2009cm}.
Thus, the charmless two-body nonleptonic $\Lambda^0_b$ decay amplitudes are determined by the hadronic matrix elements of the corresponding local four-quark operators
\begin{equation}
    \begin{aligned}
        \langle \mathcal{B}M| \mathcal{H}_{eff}| \Lambda_b^0(p_i) \rangle = \frac{G_{F}}{\sqrt{2}}V_{\rm{CKM}}\sum_i C_i\langle \mathcal{B}M| O_i | \Lambda_b^0 \rangle \ ,
    \end{aligned}
\end{equation}
where $M$ and $\mathcal{B}$ denote the meson and baryon in the final state, respectively, and $V_{\text{CKM}}$ stands for the relevant combination of CKM matrix elements. For the tree operators of the decay processes, the involved combinations of the Wilson coefficients are $a_1(\mu)=C_1(\mu)+C_2(\mu)/3=1.03$ and $a_2(\mu)=C_2(\mu)+C_1(\mu)/3=0.103$ at $\mu=m_b$, respectively~\cite{Lu:2009cm}. For penguin operators, we take $a_3(\mu)=C_3(\mu)+C_4(\mu)/3=3.6\times 10^{-3}$, $a_4(\mu)=C_3(\mu)/3+C_4(\mu)=2.3\times 10^{-2}$, $a_5(\mu) = C_5(\mu) + C_6(\mu)/3 = -2.29 \times 10^{-3}$ and $a_6(\mu) = C_5(\mu)/3 + C_6(\mu) = -29.8 \times 10^{-3}$ at $\mu=m_b$, $a_7(\mu) = C_7(\mu) + C_8(\mu)/3 = 12.2 \times 10^{-4}, a_8(\mu) = C_7(\mu)/3 + C_8(\mu) = 7.57 \times 10^{-4}, a_9(\mu) = C_9(\mu) + C_{10}(\mu)/3 = -82.2 \times 10^{-4}, a_{10}(\mu) = C_9(\mu)/3 + C_{10}(\mu) = -8.20 \times 10^{-4}$ at $\mu=m_b$~\cite{Lu:2009cm}. 

With the aid of the naive factorization hypothesis \cite{Fakirov:1977ta}, the hadronic matrix elements can be approximated as a product of two separate matrix elements
\begin{align}
\langle \mathcal{B} M | O_i | \Lambda_b^0 \rangle \to \langle M | {J}_{i,1} | 0 \rangle \otimes \langle \mathcal{B} | {J}_{i,2} | \Lambda_b^0 \rangle.
\end{align}
For instance, when $O_i=O_{1,2}$, ${J}_{i,1}$ and ${J}_{i,2}$ correspond to the {$(V-A)$} currents with the quark flavor transitions .
The matrix elements $\langle M | {J}_{i,1} | 0 \rangle$ and $\langle \mathcal{B} | {J}_{i,2} | \Lambda_b^0 \rangle$ are parametrized by the decay constants of the emitted mesons and the heavy-to-light form factors of the $\Lambda_b^0 \to \mathcal{B}$ transitions.
This factorization method allows for tractable calculations, although it neglects certain nonfactorizable effects. 
The amplitudes of charmless two-body nonleptonic $\Lambda^0_b$ decays take the following forms,
\begin{itemize}
    \item for decays involving a pseudoscalar or scalar meson in the final state,
\begin{equation}
    \begin{aligned}
        \mathcal{A}(\Lambda_b^0(p_i) \rightarrow \mathcal{B}(p_4)P/S) = i\bar{u}(p_4)[A + B\gamma_5]u(p_i)\ ;
    \end{aligned}
\end{equation}
\item for decays involving a vector meson in the final state,
\begin{equation}
    \begin{aligned}
        \mathcal{A}(\Lambda_b^0(p_i) \rightarrow \mathcal{B}(p_4)V) = \bar{u}(p_4)[A_1\gamma_{\mu}\gamma_5 + A_2\frac{p_{f,\,\mu}}{m_i}\gamma_5 + B_1\gamma_{\mu} + B_2\frac{p_{f,\,\mu}}{m_i}]\epsilon^{*\mu}u(p_i) \ ,
    \end{aligned}
\end{equation}
\end{itemize}
where $u(p_i)$ and $u(p_4)$ denote the Dirac spinors of the initial and final baryons, respectively, and $\epsilon^{*\mu}$ is the polarization of the final-state vector meson~\cite{Lee:1957qs, Pakvasa:1990if}. The invariant amplitudes $A$, $B$, $A_i$, and $B_i$ ($i=1,2$) are the combinations of the the Wilson coefficients, the decay constants, and the heavy-to-light transition form factors, which encode the dynamics of the underlying hadronic transition. Detailed definitions and phenomenological discussions of these invariant amplitudes can be found in Refs.~\cite{Zhu:2018jet, Lee:1957qs, Pakvasa:1990if}.
It is worth emphasizing that short-distance contributions not only encode the effects of high-scale physics but also play a crucial role in shaping the long-distance hadronic dynamics, as will be further illustrated in the subsequent discussion.

\subsection{Long-distance contributions within the framework of rescattering mechanism}
\begin{figure}[thb]
\begin{center}
\includegraphics[width=\columnwidth]{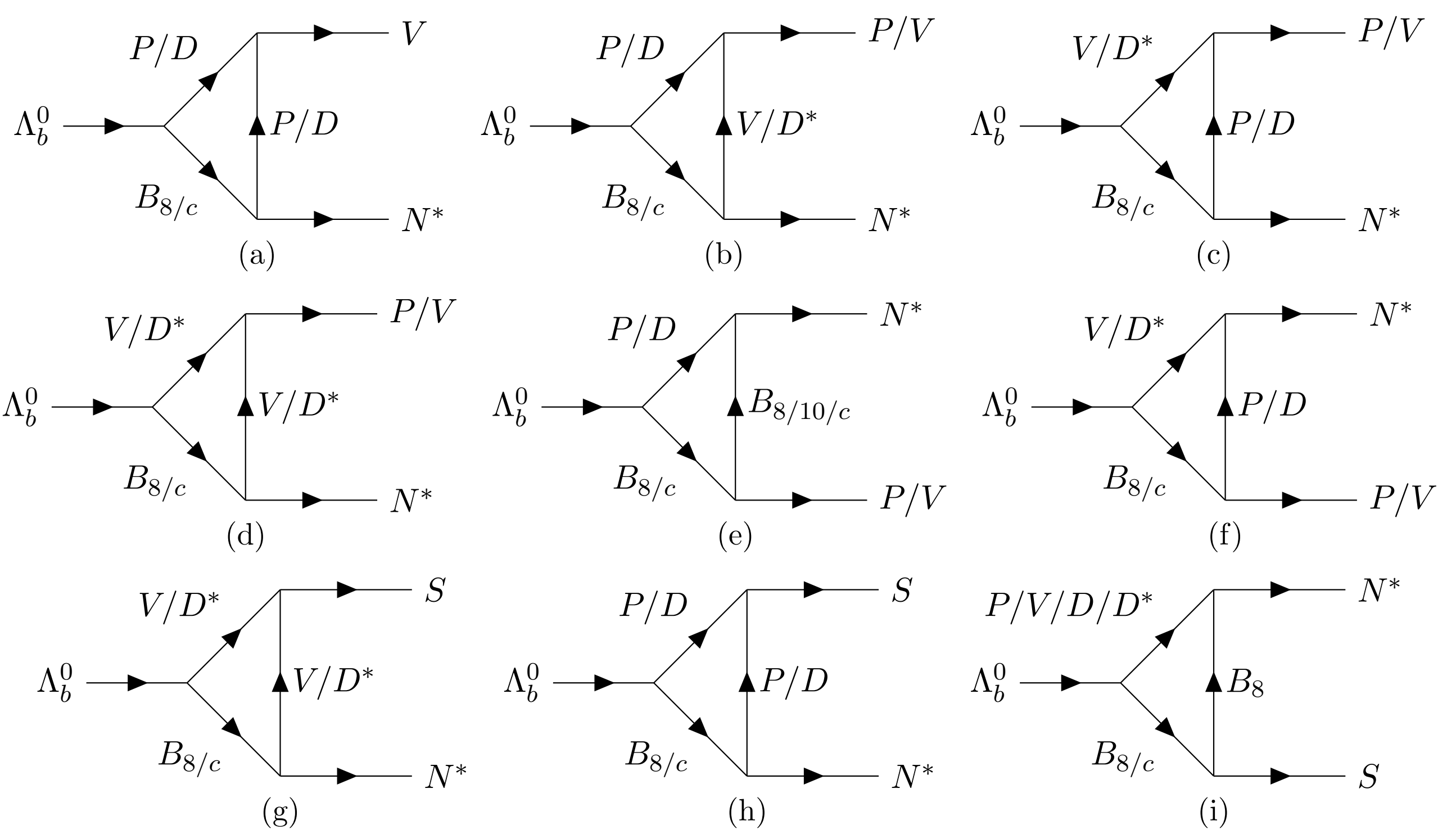}
\end{center}
\vspace{-20pt}
\caption{The diagrams of $\Lambda_b^0 \to N^* M$ with $M =K_S, K^*_0(700)$, $f_0(500,980), \rho(770), \bar{K}^{*0}$, $\phi$, from rescattering mechanism at hadronic level with exchanging one single particle, where the $B_8, B_c, B_{10}, N^*$ denote octet baryons, charmed baryons, decuplet baryons and $N^*(1520,1535)$, respectively, and $P,S,V,D$ and $D^*$ are pseudoscalar, vector, $D$ and $D^*$ mesons, respectively.}
\label{fig:Triangle diagram}
\end{figure}

In charmless two-body nonleptonic $\Lambda^0_b$ decay processes, long-distance contributions are of substantial significance, particularly for the prediction of the relative strong phases and subsequently for the $CP$ asymmetries. The final-state rescattering mechanism offers a theoretically well-founded and physically transparent framework for characterizing long-distance contributions in $\Lambda_b$ decays.
Based on $CPT$ invariance and unitarity, Wolfenstein and Suzuki proposed a formalism for final-state interactions at the hadronic level \cite{Wolfenstein:1990ks, Suzuki:1999uc}. Subsequently, Cheng, Chua, and Soni performed a systematic study of two-body B meson decays, including the effects of final-state interactions \cite{Cheng:2004ru}. In the time-evolution picture, short-distance interactions associated with weak decays occur promptly at early times, whereas long-distance rescattering effects emerge at later stages as the decay products propagate and interact. Thus, the full amplitude can be expressed as~\cite{Cheng:2020ipp}
\begin{equation}
    \begin{aligned}
        \mathcal{A}(\Lambda_b^0 \rightarrow f) = \sum_i \langle f|U(+\infty, \tau)|i \rangle \langle i | \mathcal{H}_{eff}|\Lambda_b^0 \rangle\ ,
    \end{aligned}
\end{equation}
where the index $i$ runs over all possible intermediate states to account for the complete rescattering contributions. 
The matrix element $\langle i| \mathcal{H}_{\text{eff}} |\Lambda_b^0 \rangle$ does not generate strong phases, while the rescattering part $\langle f|U(+\infty, \tau )|i\rangle$ introduces a complex amplitude with a nonzero phase. Since the long-distance contributions are intrinsically nonperturbative, they can be estimated at the hadronic level via a single-particle-exchange approximation. Thus, the long-distance contributions to the charmless two-body nonleptonic $\Lambda_b^0$ decays are modeled via hadronic triangle diagrams, which corresponds to the rescattering process between two intermediate hadrons after the weak decay of the $\Lambda_b^0$, as illustrated in Fig.~\ref{fig:Triangle diagram}. Previous studies~\cite{Magalhaes:2011sh, Garrote:2022uub} have shown that two-body scatterings dominate the final-state interactions, while the contributions from three-body and four-body intermediate states are relatively small. Therefore, we neglect these contributions in this work. 

The diagrams in Fig.~\ref{fig:Triangle diagram} effectively encode the final-state interactions and provide a natural mechanism for incorporating rescattering effects into the decay amplitudes.
To evaluate these triangle diagrams, we need to combine the weak interaction vertex, treated as a short-distance amplitude under the naive factorization hypothesis, with the strong interaction vertices of the rescattering processes. The corresponding effective Lagrangian and Feynman rules for the strong interaction vertices are collected in Appendix \ref{EL} and Appendix \ref{FR}, respectively.

\begin{figure}[thb]
\begin{center}
\includegraphics[width=0.5\columnwidth]{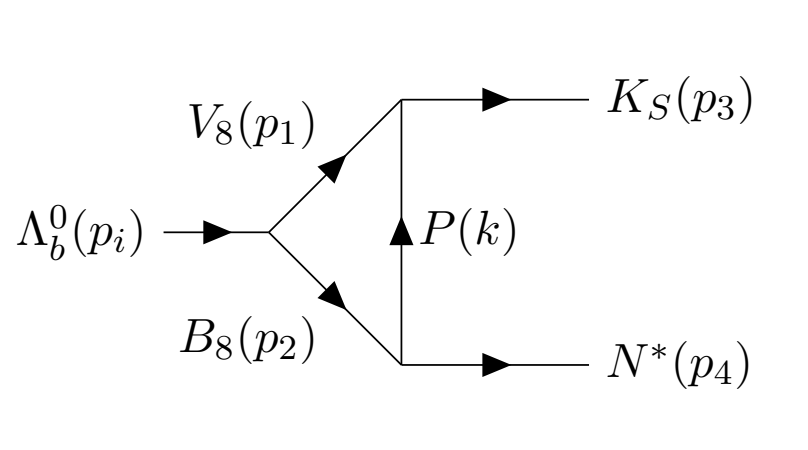}
\end{center}
\vspace{-20pt}
\caption{The diagram of $\Lambda_b^0 \to N^* K_S$ with intermediate states octet vector meson $V_8$ and octet baryons $B_8$ rescattering by exchanging a pseudoscalar meson.}
\label{fig: long distance example}
\end{figure}

Employing the established Feynman rules of strong interaction vertices and the factorization formula of weak transition amplitudes, we can derive the amplitudes for the triangle diagrams displayed in Fig.~\ref{fig:Triangle diagram} with the loop integral variable $k$, by performing the relevant hadronic loop integrals. Then we give an example of the decay $\Lambda_b^0 \to N^*  K_S$, with intermediate states an octet vector meson $V_8 (p_1)$ and an octet baryon $B_8(p_2)$ rescattering by exchanging a pseudoscalar meson $P(k)$, as shown in Fig. \ref{fig: long distance example}. The final expressions for the $\Lambda_b^0 \to N^*(1520, 1535) K_S$ amplitudes are integrals of the loop momentum $k$,
\begin{equation}
    \begin{aligned} 
		\mathscr{M}[V_8,B_8; P](1535) &= -i^6 \int \frac{d^4k}{(2\pi)^4}\frac{g_{VPP}}{\sqrt{2}}\frac{f_{\pi NN^*}^{1535}}{m_{\pi}}\bar{u}(p_4,s_4)\gamma_{\mu}k^{\mu}(\not{p}_2 + m_2)[A_1\gamma_{\nu}\gamma_5 + A_2 \frac{p_{2\nu}}{m_i}\gamma_5 \\
        & + B_1 \gamma_{\nu} + B_2\frac{p_{2\nu}}{m_i}] \times u(p_i,s_i) \times (-g^{\nu \alpha} + \frac{p_{1}^{\nu}p_{1}^{\alpha}}{m_1^2})(p_3 + k)_{\alpha} \times \mathcal{F}\times \mathcal{P} \ ,\\
        \mathscr{M}[V_8,B_8; P](1520) &= i^5 \int \frac{d^4k}{(2\pi)^4}\frac{g_{VPP}}{\sqrt{2}}\frac{f_{\pi NN^*}^{1520}}{m_{\pi}} \bar{u}_{\mu}(p_4,s_4)\gamma_5 k^{\mu}(\not{p}_2 + m_2)[A_1\gamma_{\nu}\gamma_5 + A_2 \frac{p_{2\nu}}{m_i}\gamma_5 \\ 
        & + B_1 \gamma_{\nu} + B_2\frac{p_{2\nu}}{m_i}] \times u(p_i,s_i)\times (-g^{\nu \alpha } + \frac{p_{1}^{\nu}p_{1}^{\alpha}}{m_1^2})(p_3 + k)_{\alpha} \times \mathcal{F} \times \mathcal{P} \  , 
    \end{aligned}
\end{equation}
where $f_{\pi NN^*}^{(1520, 1535)}$ and $g_{VPP}$ are the corresponding strong coupling constants and the coefficients $A_1$, $A_2$, $B_1$, and $B_2$ are transition form factors.  And $m_i$ are the mass of the initial state $\Lambda_b^0$. $A_1$, $A_2$, $B_1$, $B_2$ encode the short-distance contributions. Besides, $\mathcal{P}$ is the compact form of the product of the involved propagators,  
\begin{equation}
    \begin{aligned}
         \mathcal{P} = \frac{1}{p_1^2 - m_1^2 + i\epsilon} \cdot \frac{1}{p_2^2 - m_2^2 + i\epsilon} \cdot \frac{1}{k^2 - m_k^2 + i\epsilon} \ ,
    \end{aligned} \label{eq: propagator}
\end{equation}
where $p_1,p_2,k$ are the momenta of the three intermediate particles as shown in Fig.~\ref{fig: long distance example}, and $m_1,m_2,m_k$ are their masses, while $\mathcal{F}$ is the regulator form factors for the exchange particles. The complete set of explicit expressions for all the triangle diagrams computed in this work are provided as the Supplemental Material~\cite{SM} and the full amplitudes of $\Lambda_b \to N^* M$ with $M =K_S, K^*_0(700)$, $f_0(500,980), \rho(770), \bar{K}^{*0}$, $\phi$, are collected in Appendix~\ref{FA}.
Following Ref.~\cite{Duan:2024zjv}, we adopt the following regulator form factor to describe the off-shell effects of the intermediate hadrons and regulate the ultraviolet behavior of the loop integrals~\cite{Yue:2024paz}

\begin{equation}
    \begin{aligned}
        \mathcal{F}(\Lambda, k, m_k) = \frac{\Lambda^4}{(k^2 - m_k^2)^2 + \Lambda^4} \ ,  
    \end{aligned} \label{eq:form factors}
\end{equation}
where $\Lambda$ is a phenomenological model parameter. Following the Ref.~\cite{Duan:2024zjv}, we use $\Lambda_{\rm charm}$ for triangle diagrams involving charmed intermediate hadrons and $\Lambda_{\rm charmless}$ for diagrams containing only charmless intermediate hadrons. This distinction concerns the intermediate states; all final-state decay modes considered here are charmless. For example, the $\Lambda_b^0\to N^*(1535)K_S$ amplitude receives contributions from the intermediate states $\{p,K^{*-}\}$ with $\pi^+$ exchange and $\{\Lambda_c^+,D_s^{*-}\}$ with $D^-$ exchange. These contributions are evaluated using $\Lambda_{\rm charmless}$ and $\Lambda_{\rm charm}$, respectively. We adopt $\Lambda_{\rm charm}=1.0\pm0.1$ and $\Lambda_{\rm charmless}=0.5\pm0.1$ from Ref.~\cite{Duan:2024zjv}, where the parameters were constrained using the measured branching ratios and direct$CP$ asymmetries of $\Lambda_b^0\to p\pi^-$ and $pK^-$. They are not refitted to the $N^*M$ channels. At fixed nonzero virtuality $(k^2-m_k^2)$, increasing $\Lambda$ increases the regulator factor and weakens the off-shell
suppression, whereas decreasing $\Lambda$ strengthens it. The dependence of physical observables must nevertheless be determined from the full amplitudes and their interference.

\section{NUMERICAL RESULTS AND ANALYSIS}
In this section, we summarize the input parameters used for numerical calculations and present the results for the branching ratios, the asymmetry parameters, and the $CP$ asymmetries of the following decay channels, categorized by the type of final state mesons:
\begin{itemize}
    \item Pseudoscalar meson: $\Lambda_b^0 \to N^*(1535,1520)K_S^0$
    \item Scalar meson: $\Lambda_b^0 \to N^*(1535,1520)K^{*}_{0}(700)/f_0(500, 980)$
    \item Vector meson: $\Lambda_b^0 \to N^*(1535,1520)\rho^0/\bar{K}^{*0}/\phi$.
\end{itemize}

\subsection{Basic inputs}
In this work, all the masses of quarks and hadrons used as input parameters are taken from Ref.~\cite{ParticleDataGroup:2022pth} and listed in Table.~\ref{tab:masses_all}. The light quark masses correspond to the $\overline{\text{MS}}$ masses evaluated at the renormalization scale $\mu = 2~\text{GeV}$, and the heavy quark masses $m_b$ and $m_c$ are given in the $\overline{\text{MS}}$ scheme at the scale of their respective $\overline{\text{MS}}$ masses. The decay constants of the pseudoscalar and vector mesons are taken from Ref.~\cite{Zhu:2018jet} and summarized in Table.~\ref{tab:decay_constants}. The CKM matrix is expanded in the Wolfenstein parametrization as~\cite{Kobayashi:1973fv, Cabibbo:1963yz}
\begin{align}
V_{\text{CKM}} =
\begin{pmatrix}
V_{ud} & V_{us} & V_{ub} \\
V_{cd} & V_{cs} & V_{cb} \\
V_{td} & V_{ts} & V_{tb}
\end{pmatrix} = \begin{pmatrix}
    1 - \lambda_W^2/2 & \lambda_W & A\lambda_W^3(\rho - i\eta) \\ 
    -\lambda_W & 1 - \lambda_W^2/2 & A\lambda_W^2 \\
    A\lambda_W^3(1 - \rho -i\eta) & -A\lambda_W^2 & 1
\end{pmatrix} + \mathcal{O}(\lambda_W^4),
\end{align}
where the Wolfenstein parameters are taken as $A = 0.823$, $\rho = 0.141$, $\eta = 0.349$, and $\lambda_W = 0.225$. Here the subscript $W$ is introduced to the parameter $\lambda_W$ to distinguish it from the $\lambda$ function used in Ref.~\cite{ParticleDataGroup:2022pth}.
\begin{table}[htbp]
\centering
\caption{Masses of baryons, mesons, and quarks used in this work~\cite{ParticleDataGroup:2022pth}.}
\label{tab:masses_all}
\renewcommand{\arraystretch}{1.4}
\setlength{\tabcolsep}{4pt}
\begin{tabular}{lccccccc}
\hline\hline
\textbf{Parameters} & $\Lambda_b$ & $\Lambda_c^+$ & $p$ & $n$ & $N^*(1520)$ & $N^*(1535)$ & $\Lambda^0$ \\
\textbf{Mass (GeV)} & 5.619 & 2.286 & 0.938 & 0.940 & 1.515 & 1.530 & 1.116 \\
\hline
\textbf{Parameters} & $D^{\pm}$ & $D^0$ & $D^*$ & $D_s$ & $D_s^*$ & $K$ & $K^*$ \\
\textbf{Mass (GeV)} & 1.869 & 1.865 & 2.007 & 1.968 & 2.106 & 0.493 & 0.892 \\
\hline
\textbf{Parameters} & $\rho$ & $\pi$ & $\omega$ & $\phi$ & $f_0(500)$ & $f_0(980)$ &  \\
\textbf{Mass (GeV)} & 0.770 & 0.140 & 0.783 & 1.020 & 0.513 & 0.980 &  \\
\hline
\textbf{Parameters} & $u$ & $d$ & $s$ & $c$ & $b$ &  &  \\
\textbf{Mass (MeV)} & 2.16 & 4.70 & 93.5 & 1270 & 4180 &  &  \\
\hline\hline
\end{tabular}
\end{table}

\begin{table}[htbp]
\centering
\caption{Decay constants of pseudoscalar and vector mesons used in this work~\cite{Zhu:2018jet}.}
\label{tab:decay_constants}
\renewcommand{\arraystretch}{1.3}
\setlength{\tabcolsep}{6pt}
\begin{tabular}{ccccccccccc}
\hline\hline
\textbf{Decay constant} & $f_{\pi}$ & $f_{K}$ & $f_{\rho}$ & $f_{\omega}$ & $f_{\phi}$ & $f_{K^*}$ & $f_{D^*}$ & $f_{D}$ & $f_{D_s}$ & $f_{D_s^*}$ \\
\textbf{Value [MeV]} & 130.3 & 155.5 & 216 & 187 & 215 & 210 & 230 & 212 & 250 & 271 \\
\hline
\textbf{Decay constant} & $f_{J/\psi}$ & $f_{\eta_c}$ & $f_{\eta_u}$ & $f_{\eta_d}$ & $f_{\eta_s}$ & $f_{\eta_u'}$ & $f_{\eta_d'}$ & $f_{\eta_s'}$ &  &  \\
\textbf{Value [MeV]} & 418 & 387 & 54 & 54 & $-111$ & 44 & 44 & 136 &  &  \\
\hline\hline
\end{tabular}
\end{table}

In calculation of the short distance contributions, the weak transition $\Lambda_b \to p, n, \Lambda^0, \Lambda_c$ form factors serve as the essential hadronic input parameters and have been investigated within various QCD-based theoretical methods, such as the perturbative QCD approach~\cite{Lu:2009cm}, (covariant) light-front quark model~\cite{Ke:2007tg,Wei:2009np}, QCD factorization~\cite{Zhu:2016bra}, soft-collinear effective theory~\cite{Feldmann:2011xf} and light-cone sum rules~\cite{Khodjamirian_2011}.
In this work, we employ the theoretical results for the heavy-to-light baryonic form factors obtained in Ref. \cite{Zhu:2018jet}, which are summarized in Table.~\ref{tab:form factors}.

In the long distance contribution, the strong coupling constants are the fundamental nonperturbative inputs. We adopt the values derived from light-cone sum rules, which take into account the $\rm{SU(3)}_f$ symmetry breaking effects \cite{Aliev:2006xr, Aliev:2009ei}. For the strong coupling constants involving the excited nucleon states $N^*(1535, 1520)$, the values are taken from Ref.~\cite{Riska:2000gd}. The strong coupling constants for the excited nucleon states $N^*(1535, 1520)$ and charmless mesons are
\begin{equation}\label{eq:Ncoupling}
f_{\pi NN^*}^{1520} = -1.71, 
f_{\pi NN^*}^{1535} = 0.49, 
g_{\omega NN^*}^{1520} = 7.7, 
g_{\omega NN^*}^{1535} = -4.5, 
g_{\rho NN^*}^{1520} = 4.5,
g_{\rho NN^*}^{1535} = -2.9 \ .
\end{equation}

\begin{table}[htbp]
    \centering
    \caption{The values of the $\Lambda_b \to p, n, \Lambda^0, \Lambda_c$ transition form factors from covariant light-front quark model~\cite{Zhu:2018jet}. The definitions of $f_{1,2}$ and $g_{1,2}$ can be found in~\cite{Zhu:2018jet}.}
    \label{tab:form factors}
    \renewcommand{\arraystretch}{1.2} 
    \setlength{\tabcolsep}{28pt}      
    \begin{tabular}{lccccccccccc} \\ \hline \hline
       Decay & $f_1(0)$ & $f_2(0)$ & $g_1(0)$ & $g_2(0)$ \\ \hline
       $\Lambda_b \to \Lambda_c^+$ & $0.500$ & $-0.098$ & $0.509$ & $-0.015$ \\ \hline 
       $\Lambda_b \to p$ & $0.128$ & $-0.056$ & $0.129$ & $-0.033$ \\ \hline 
       $\Lambda_b \to \Lambda^0$ & $0.131$ & $-0.048$ & $0.132$ & $-0.023$ \\ \hline 
       $\Lambda_b \to n$ & $0.128$ & $-0.056$ & $0.129$ & $-0.033$ \\ \hline 
    \end{tabular}
\end{table}

The remaining strong coupling constants of this type involving charmed mesons are determined under the assumption of SU(4) flavor symmetry \cite{yue2024strongdecayslambdac2910lambdac2940}. Furthermore, the strong couplings for two charmed mesons and one light meson are taken from Ref.~\cite{Cheng:2004ru}. The values are given by
\begin{equation}
g_{D^*D^*P_8} = \frac{g_{D^*DP_8}}{\sqrt{m_D m_{D^*}}}, \qquad  
g_{DDV} = g_{D^*D^*V} = \frac{\beta g_V}{\sqrt{2}}, \qquad  
f_{D^*DV} = \frac{f_{D^*D^*V}}{m_{D^*}} = \frac{\lambda g_V}{\sqrt{2}} \, ,
\end{equation}
where $g_V = m_\rho/f_{\pi}$, $\beta = 0.9$, $\lambda = 0.56~\text{GeV}^{-1}$ and $g_{D^*DP_8}=17.9$. $P_8$ denotes the octet pseudoscalar meson and $g_{D^*DP_8}$ is extracted from the experimental data of the $D^*$ decay width \cite{Cheng:2004ru}. In addition, the strong coupling constants for charmed hadrons and light baryon octet are taken from Ref.~\cite{Yu:2017zst}. These strong coupling constants are defined in the corresponding effective Lagrangians collected in Appendix~\ref{EL}.

\subsection{Branching ratios}
We first calculate the decay widths of the decay channels. The expression for the decay width is obtained by summing the squares of all the helicity amplitudes over the spins of the initial and final states, and averaging over the spin of the initial particle, as given by:
\begin{align}
\Gamma(\Lambda_b \to N^*M) = \frac{p_c}{8\pi m_{\Lambda_b}^2} \frac{1}{2} \sum_{\lambda_i \lambda_f \lambda_M} |H^{\lambda_i}_{\lambda_f, \lambda_M}(\Lambda_b \to N^*M)|^2,
\end{align}
where $p_c=\frac{1}{2m_{\Lambda_b}}\sqrt{[m^2_{\Lambda_b}-(m^2_{N^*}+m^2_M)][m^2_{\Lambda_b}-(m^2_{N^*}-m^2_M)]}$ is the momentum of $N^*$ in the rest frame of $\Lambda_b$, and 
$H$ denotes the helicity amplitudes with the indices $\lambda_i$, $\lambda_f$, and $\lambda_M$, labeling the helicities of the initial-state baryon, the final-state baryon and the meson, respectively. Accordingly, the branching ratios are obtained by dividing the decay width of each channel by the total decay width of $\Lambda_b$. 

Based on the above formulas, the input parameters, and the model parameters $\Lambda_{\mathrm{charm}} = 1.0 \pm 0.1$ and $\Lambda_{\mathrm{charmless}} = 0.5 \pm 0.1$, the $CP$-averaged $\Lambda_b \to N^*M$ branching ratios are evaluated, and the results are shown in Table.~\ref{tab:Br_summary}. The uncertainties arise from the uncertainties of the introduced model parameters. We do not consider the uncertainties induced by the variation of the scale $\mu$, because these uncertainties can be safely neglected in comparison with those induced by the model parameters.

From Table.~\ref{tab:Br_summary}, we find that the branching ratio of the $\Lambda_b \to N^*(1535)K_S$, with the pseudoscalar meson in the final state, is of $\mathcal{O}(10^{-5})$, while the $\Lambda_b \to N^*(1535)K^{*}_{0}(700)$ with the scalar meson in the final state is of $\mathcal{O}(10^{-8})$. Both of them are dominated by charm-loop triangle diagrams, but involve different charmed hadrons in the loops, as $K_S$ and $K_0^*(700)$ have different parities. For $\Lambda_b \to N^*(1535)K_S$, the loop contains the $D^*\Lambda_c N^*$ vertex; while for $\Lambda_b \to N^*(1535)K^{*}_{0}(700)$, the loop contains the $D\Lambda_c N^*$ vertex. From the corresponding Lagrangians in Eq.~\eqref{eq:LN*} and the values for the couplings in Eq.~\eqref{eq:Ncoupling}, it can be seen that the magnitude of the former is enhanced by the factors of $m_{\Lambda_b}/m_D\sim 3$ and ${g_{VNN^*}}/{g_{PNN^*}}\sim 6$, and can thus account for the branching ratio hierarchy of nearly three orders of magnitude. In addition, as the $SVV$ couplings are not available currently, we have neglected the involved contributions to $\Lambda_b \to N^*K^{*}_{0}(700)$. 
Therefore, $\Lambda_b \to N^*K^{*}_{0}(700)$  and also the  $\Lambda_b \to N^*f_0(500,980)$ channels, in which the final states involve a scalar meson, await the couplings to have more reliable predictions.
Among the decay channels with a vector meson in the final state, the branching ratio of $\Lambda_b \to N^*(1535)\phi$ is extremely small, which is due to the Okubo-Zweig-Iizuka suppression~\cite{OKUBO1963165}. The partial-wave branching ratios further show that $\Lambda_b\to N^*(1535)K_S$ is dominated by the $P$-wave contribution, whereas $\Lambda_b\to N^*(1520)K_S$ and most scalar-meson modes are dominated by the $S$-wave contribution. In addition to these features of the central values, the uncertainties quoted in Table.~\ref{tab:Br_summary} indicate that the absolute branching
ratios are sensitive to variations of the two model parameters. In many channels, the upper uncertainties are larger than the lower ones, which is consistent with the behavior of the regulator form factor: increasing the model parameters weakens the off-shell suppression and enhances the loop amplitudes.
\begin{table}[htbp]
\centering
\caption{Branching ratios and partial-wave branching ratios of $\Lambda_b \to N^*M$.}
\label{tab:Br_summary}
\renewcommand{\arraystretch}{1.4}
\setlength{\tabcolsep}{10pt}
\begin{tabular}{lccc}
\hline\hline
\textbf{Decay channel}  & \textbf{Br} & $\textbf{Br}^{S-\rm{wave}}$ & $\textbf{Br}^{P-\rm{wave}}$\\
\hline
$N^*(1535)K_S$ & $(1.62^{+1.79}_{-0.91}) \times 10^{-5}$ 
& $(0.10^{+0.11}_{-0.06})\times 10^{-5}$
& $(1.52^{+1.69}_{-0.85})\times 10^{-5}$ \\
$N^*(1535)K^*_0(700)$ & $(3.16^{+0.42}_{-0.22})\times 10^{-8}$
& $(2.40^{+0.32}_{-0.17})\times 10^{-8}$
& $(0.76^{+0.33}_{-0.20})\times 10^{-8}$ \\
$N^*(1535)f_0(500)$ & $(4.18^{+3.81}_{-1.91}) \times 10^{-7}$
& $(3.03^{+2.74}_{-1.40})\times 10^{-7}$
& $(1.14^{+1.07}_{-0.51})\times 10^{-7}$ \\
$N^*(1535)f_0(980)$ & $(1.55^{+1.76}_{-0.87})\times 10^{-6}$ 
& $(1.43^{+1.61}_{-0.81})\times 10^{-6}$
& $(0.12^{+0.15}_{-0.07}) \times 10^{-6}$\\
\hline
$N^*(1520)K_S$ & $(7.17^{+8.01}_{-4.05})\times 10^{-5}$
& $(6.55^{+7.31}_{-3.69})\times 10^{-5}$
& $(0.62^{+0.70}_{-0.36})\times 10^{-5}$ \\
$N^*(1520)K^*_0(700)$ & $(1.92^{+1.67}_{-0.90})\times 10^{-7}$ 
& $(1.91^{+1.64}_{-0.90})\times 10^{-7}$
& $1.65^{+3.32}_{-0.90})\times 10^{-9}$\\
$N^*(1520)f_0(500)$ & $(6.09^{+1.95}_{-1.10})\times 10^{-8}$
& $(4.18^{+1.08}_{-0.73})\times 10^{-8}$
& $(1.91^{+0.91}_{-0.51})\times 10^{-8}$ \\
$N^*(1520)f_0(980)$ & $(4.59^{+4.60}_{-2.37})\times 10^{-7}$ 
& $(3.18^{+3.14}_{-1.62})\times 10^{-7}$
& $(1.40^{+1.46}_{-0.74})\times 10^{-7}$\\
\hline
$N^*(1535)\bar{K}^{*0}$ & $(1.66^{+1.90}_{-0.95}) \times 10^{-4}$
& $...$
& $...$ \\
$N^*(1535)\rho(770)$ & $(0.85^{+0.56}_{-0.33})\times 10^{-4}$
& $...$
& $...$ \\
$ N^*(1535)\phi$ & $(1.89^{+5.69}_{-1.55})\times 10^{-10}$
& $...$
& $...$ \\ \hline
$ N^*(1520)\bar{K}^{*0}$ & $(1.58^{+1.69}_{-0.86})\times 10^{-4}$
& $...$
& $...$ \\
$ N^*(1520)\rho(770)$ & $(1.44^{+1.19}_{-0.61})\times 10^{-4}$ 
& $...$
& $...$\\
$ N^*(1520)\phi$& $(1.05^{+3.40}_{-0.87})\times 10^{-9}$
& $...$
& $...$ \\
\hline\hline
\end{tabular}
\end{table}

\begin{figure}[thb]
\begin{center}
\includegraphics[width=\columnwidth]{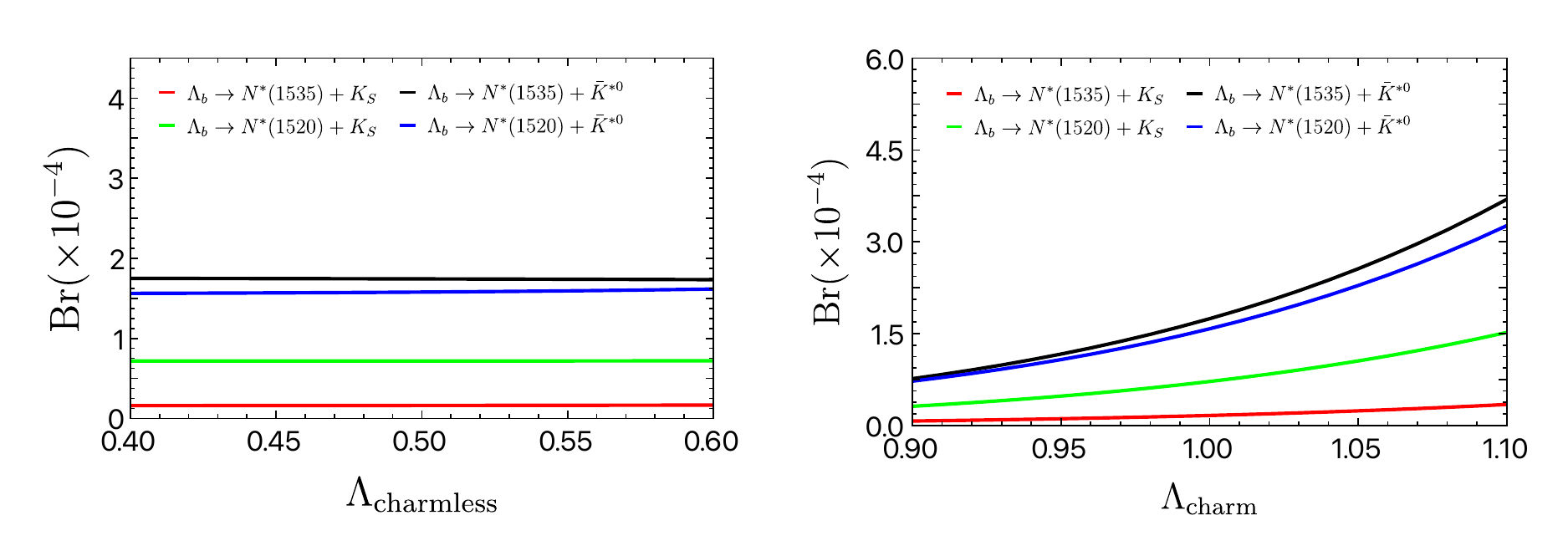}
\end{center}
\vspace{-20pt}
\caption{The dependence of the average branching ratios of $\Lambda_b \to N^*(1535, 1520)K_S/\bar{K}^{*0}$ on the model parameters $\Lambda_{\rm charmless}$ and $\Lambda_{\rm charm}$.}
\label{fig:Br}
\end{figure}

Figure ~\ref{fig:Br} shows the dependence of the branching ratios on the two model parameters for the $N^*(1535,1520)K_S/\bar{K}^{*0}$ decay channels. Over the ranges considered, these branching ratios are much more sensitive to $\Lambda_{\rm charm}$ than to $\Lambda_{\rm charmless}$. This behavior reflects the dominance of the charmed rescattering contributions in these channels. Variations of the dominant amplitudes strongly affect the total rates, whereas variations of the subleading charmless contributions have a smaller effect.

Apart from the model parameters, the decay branching ratio still depends on two additional input parameters, namely the heavy-to-light form factors and the strong coupling constants. 
These two parameters can be calculated using QCD-based theoretical methods or extracted from experimental data. In particular, for the couplings involving an excited nucleon $N^*$, a charmed baryon $\Lambda_c$, and a charmed meson $D$, we resort to SU(4) flavor symmetry to estimate the corresponding coupling constants \cite{Yue:2024paz}, which will introduce additional sources of uncertainty. Thus, improving the precision of these strong couplings is essential for more reliable theoretical predictions and for a deeper understanding of the dynamics of charmless nonleptonic $\Lambda_b$ decays.

\subsection{$CP$ asymmetries}
We are now ready to explore the $CP$ asymmetries of the $\Lambda_b \to N^*M$ channels. The global direct $CP$ asymmetry for each channel is defined by
\begin{equation}
    \begin{aligned}
        a_{CP}^{\rm dir} &= \frac{\Gamma - \bar{\Gamma}}{\Gamma + \bar{\Gamma}} ,
    \end{aligned}
\end{equation}
where $\bar{\Gamma}$ denotes the decay width of the corresponding $CP$-conjugate process $\bar \Lambda_b \to \bar N^*\bar M$.
For understanding the contributions to the $CP$ violation from different partial waves, we derive the partial wave amplitudes from helicity amplitudes for the $\Lambda_b \to N^*P/S$ processes by~\cite{Chen:2019hqi}
\begin{equation}
    \begin{aligned}
        H_{+1/2} = \frac{1}{\sqrt{2}}(S + P)\ ,\qquad  H_{-1/2} = \frac{1}{\sqrt{2}}(S - P)\ .
    \end{aligned}
\end{equation}
Subsequently, we define the partial-wave $CP$-violating observables in $\Lambda_b \to N^*P/S$ processes as \cite{Roy:2019cky}
\begin{equation}
    \begin{aligned}
        a_{CP}^S &= \frac{\lvert S \rvert^2 - \lvert \bar{S}\rvert^2}{\lvert S \rvert^2 + \lvert \bar{S}\rvert^2}\ , \qquad
        a_{CP}^P = \frac{\lvert P \rvert^2 - \lvert \bar{P}\rvert^2}{\lvert P \rvert^2 + \lvert \bar{P}\rvert^2}\ ,
    \end{aligned}
\end{equation}
and the direct $CP$ asymmetry can be expressed in terms of partial wave amplitudes as
\begin{equation}
    \begin{aligned}
        a_{CP}^{\rm dir} &
        = \frac{\lvert S \rvert^2 - \lvert \bar{S}\rvert^2 + \lvert P \rvert^2 - \lvert \bar{P}\rvert^2}{\lvert S \rvert^2 + \lvert \bar{S}\rvert^2 + \lvert P \rvert^2 + \lvert \bar{P}\rvert^2}\ .
    \end{aligned}
\end{equation}
This formula indicates that the global direct $CP$ asymmetry can be regarded as a weighted combination of $a_{CP}^S$ and $a_{CP}^P$. Consequently, if there is a cancellation between different partial waves, the total direct $CP$ asymmetry will be relatively suppressed. In Table.~\ref{tab:CP_complete}, we present the results of the direct $CP$ asymmetries and partial wave $CP$ asymmetries with the model parameters $\Lambda_{\mathrm{charm}} = 1.0 \pm 0.1$ and $\Lambda_{\mathrm{charmless}} = 0.5 \pm 0.1$.

A basic analysis of the CKM matrix elements and Wilson coefficients of the amplitudes can account for the smallness of $CP$ violation in certain decay channels, shown in Table.~\ref{tab:CP_complete}. Taking $\Lambda_b \to N^*K_S$ for instance, the dominant amplitudes of the tree-level triangle diagrams are proportional to $V_{cb}V_{cs}^*a_1$, which is of the order of $\lambda^2 a_1$. In contrast, the dominant amplitudes of the penguin-level triangle diagrams are proportional to $V_{tb}V_{ts}^*a_4$, which is of the order of $\lambda^2 a_4$. Therefore, the direct $CP$ asymmetry is found to be of the order of $a_4/a_1\sim 1\%$, in agreement with our calculation, and the analogous analysis can be applied to the remaining decay channels. 
\begin{table}[htbp]
\centering
\caption{Direct $CP$ and partial-wave $CP$ asymmetries for $\Lambda_b \to N^*P/S$ .}
\label{tab:CP_complete}
\renewcommand{\arraystretch}{1.4}
\setlength{\tabcolsep}{10pt}
\begin{tabular}{llll}
\hline\hline
\textbf{Decay channel} & {$\boldsymbol{a_{CP}^{dir}}$ (\%)} & {$\boldsymbol{a_{CP}^S}$ (\%)} & {$\boldsymbol{a_{CP}^P}$ (\%)} \\
\hline
$N^*(1535)K_S$ & ${0.23^{+0.29}_{-0.39}}$ & ${1.39^{+0.78}_{-0.97}}$ & ${0.15^{+0.27}_{-0.36}}$\\
$ N^*(1535)K^*_0(700)$ & ${13.2^{+4.50}_{-4.00}}$ & ${17.93^{+1.20}_{-1.10}}$ & ${-14.4^{+19.0}_{-15.4}}$ \\
$ N^*(1535)f_0(500)$ & ${-64.6^{+16.6}_{-15.4}}$ & ${-57.6^{+15.5}_{-15.3}}$ & ${-83.3^{+20.7}_{-15.8}}$ \\
$ N^*(1535)f_0(980)$ & ${-42.2^{+15.7}_{-21.5}}$ & ${-42.4^{+14.6}_{-20.2}}$ & ${-39.7^{+37.3}_{-46.5}}$ \\
\hline
$ N^*(1520)K_S$ & ${-1.89^{+1.26}_{-2.12}}$ & ${-1.66^{+1.10}_{-1.83}}$ & ${-4.42^{+3.00}_{-5.49}}$ \\
$ N^*(1520)K^*_0(700)$ & ${10.30^{+3.70}_{-3.50}}$ & ${9.80^{+3.60}_{-3.40}}$ & ${60.80^{+19.1}_{-33.5}}$ \\
$ N^*(1520)f_0(500)$ & ${-36.6^{+12.8}_{-7.2}}$ & ${-41.4^{+9.2}_{-0.7}}$ & ${-26.1^{+29.2}_{-21.3}}$ \\
$ N^*(1520)f_0(980)$ & ${-26.6^{+9.1}_{-11.8}}$ & ${-25.5^{+11.0}_{-14.6}}$ & ${-29.2^{+9.3}_{-12.2}}$ \\
\hline
$ N^*(1535)\bar{K}^{*0}$ & ${0.77^{+0.28}_{-0.23}}$ & $...$ & $...$ \\
$ N^*(1535)\rho(770)$ & ${12.30^{+6.10}_{-4.60}}$ & $...$ & $...$ \\
$ N^*(1535)\phi$ & $...$ & $...$ & $...$ \\
\hline
$ N^*(1520)\bar{K}^{*0}$ & ${2.66^{+3.09}_{-1.87}}$ & $...$ & $...$ \\
$ N^*(1520)\rho(770)$ & ${-46.7^{+13.3}_{-13.6}}$ & $...$ & $...$ \\
$ N^*(1520)\phi$ & $...$ & $...$ & $...$ \\
\hline\hline
\end{tabular}
\end{table}
In Table.~\ref{tab:CP_complete}, the direct $CP$ asymmetry of $\Lambda_b\to N^*(1535,1520)\bar{K}^{*0}$ lie around $(1-3)\%$. This is consistent with the regional $CP$ asymmetry $A_{CP}[\Lambda_b\to R(p\pi^-)R(K^-\pi^+)] \allowbreak = (2.7\pm0.8\pm0.1)\%$ reported in the LHCb discovery paper~\cite{LHCb:2025ray}, although they are not strictly the same physical quantities. 
In the $\Lambda_b\to N^* f_0(500,980)$ processes,  the $f_0(500,980)$ are scalar mesons with quantum numbers $I^G(J^P) = 0^+(0^{++})$. Their internal structure is believed to go well beyond a simple quark-antiquark ($q\bar{q}$) configuration. This conventional interpretation is disfavored by the mass pattern, large-$N_c$ arguments, and various phenomenological studies \cite{Giacosa:2009qh}, leading to its common interpretation as either a compact tetraquark or a $K\bar{K}$ bound state~\cite{ParticleDataGroup:2024cfk}. However, since the possibility of a $q\bar{q}$ composition has not been completely ruled out, we treat it as a conventional quark-antiquark state in this work. Based on this assumption, our theoretical results provide useful insights into the investigation of its internal structure. It should be noted that the contributions from certain triangle diagrams are temporarily neglected in our calculation, due to the lack of the corresponding coupling constants. Future studies will enable a more precise treatment of these contributions.
For the decay channel $\Lambda_b \to N^*(1535)K^*_0(700)$, the global direct $CP$ asymmetry $a_{CP}^{\mathrm{dir}}$ is about $13.2\%$, whereas $a_{CP}^S, ~a_{CP}^P$ reaches up to $17.93\%$ and $-14.4\%$, respectively, indicating a mild cancellation between the $S$- and $P$-wave contributions. Such a phenomenon was first observed in Ref.~\cite{Han:2024kgz}, where a perturbative QCD calculation was performed for $\Lambda_b \to p\pi^-/K^-$ decays. 
For the decay channel $\Lambda_b \to N^*(1520)K^*_0(700)$, although the $P$-wave $CP$ asymmetry $a_{CP}^P$ reaches up to $\sim60\%$, the global direct $CP$ asymmetry $a_{CP}^{dir}$ is only $\sim10\%$, staying close to the $S$-wave $CP$ asymmetry of this channel. This phenomenon originates from the $S$-wave dominance of the branching ratio of this channel. As one can see from Table.~\ref{tab:Br_summary}, the $S$-wave branching ratio is larger than the $P$-wave one by two orders of magnitude. 
For the decay channel $\Lambda_b \to N^* \phi$, no $CP$ violation is expected,  as only purely penguin amplitudes contribute. Consequently, neither a strong nor a weak phase difference arises. 

\begin{figure}[thb]
\begin{center}
\includegraphics[width=\columnwidth]{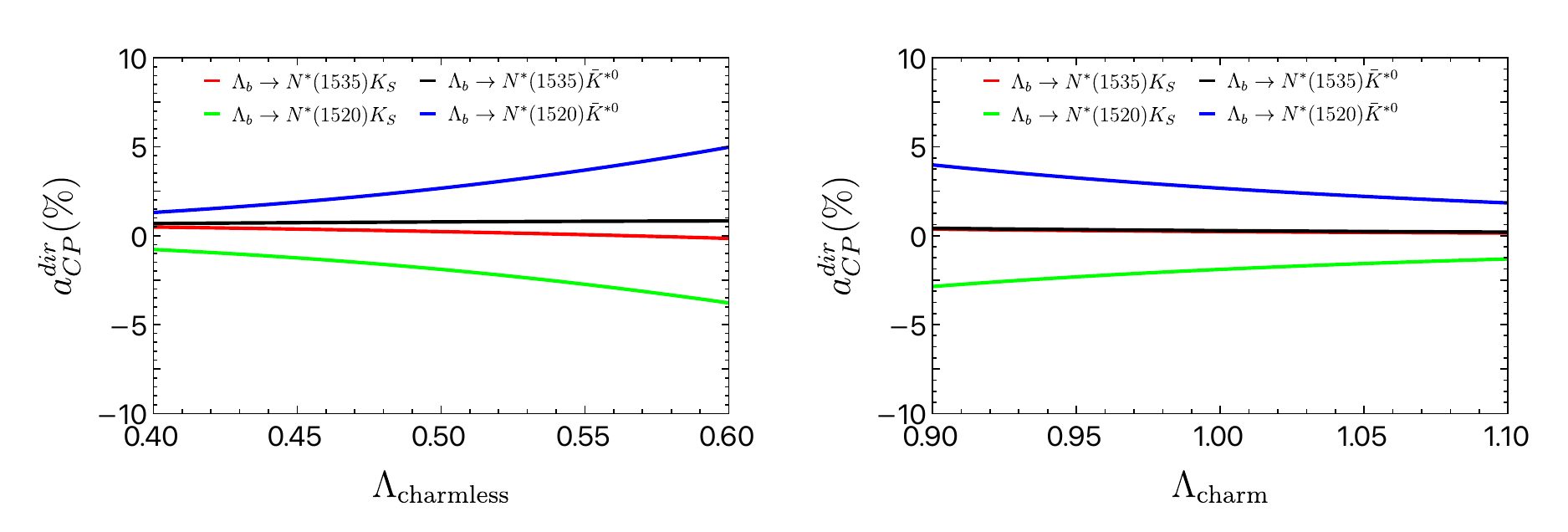}
\end{center}
\vspace{-20pt}
\caption{The dependence of the direct $CP$ asymmetries for $\Lambda_b \to N^*(1535, 1520) K_S/\bar{K}^{*0}$ on the model parameters $\Lambda_{\rm charmless}$ and $\Lambda_{\rm charm}$.}
\label{fig:directCP}
\end{figure}

\begin{figure}[thb]
\begin{center}
\includegraphics[width=\columnwidth]{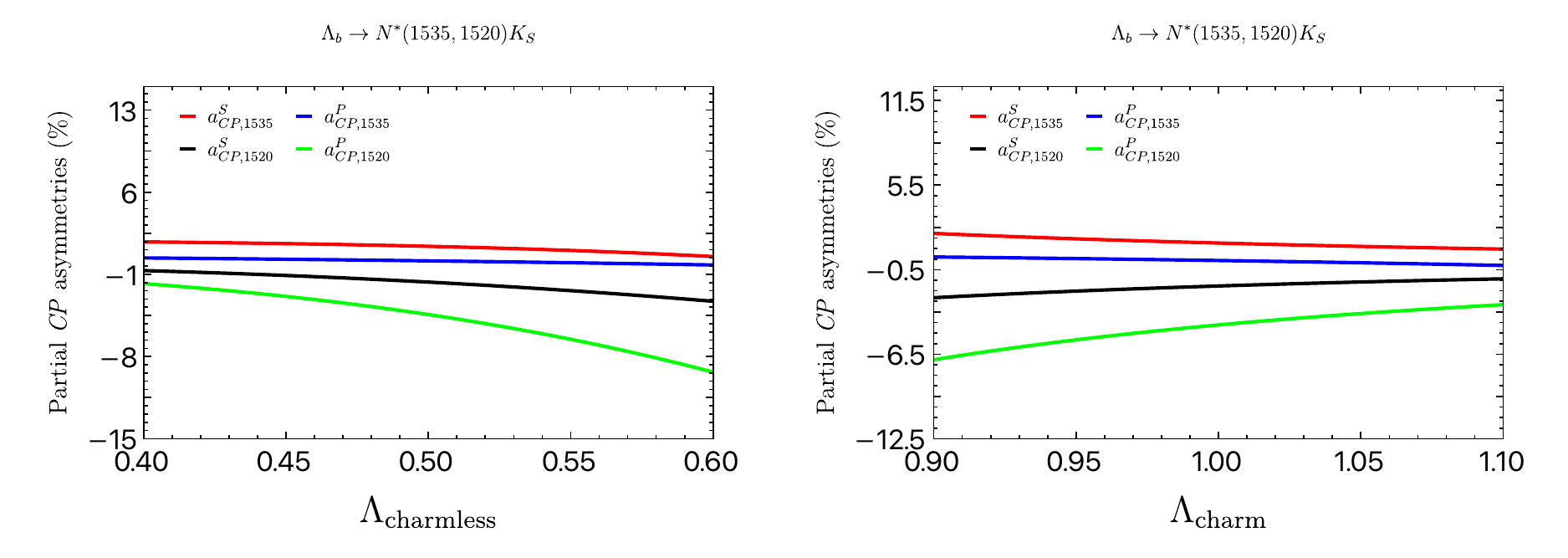}
\end{center}
\vspace{-20pt}
\caption{The dependence of the partial wave  $CP$ asymmetries for $\Lambda_b \to N^*(1535, 1520)K_S$ on the model parameters $\Lambda_{\rm charmless}$ and $\Lambda_{\rm charm}$.}
\label{fig:partialCP}
\end{figure}

In Figs.~\ref{fig:directCP} and~\ref{fig:partialCP},  we present the variations of the direct $CP$ asymmetries for $\Lambda_b \to N^*(1535, 1520) K_S/\bar{K}^{*0}$ and the partial wave $CP$ asymmetries for $\Lambda_b \to N^*(1535, 1520)K_S$ with respect to the model parameters $\Lambda_{\rm charmless}$ and $\Lambda_{\rm charm}$. In contrast to the charm decays~\cite{Jia:2024pyb},  where the predicted $CP$ asymmetries are basically independent of the model parameter, several of the observables considered here exhibit a visible dependence on both $\Lambda_{\rm{charm}}$ and $\Lambda_{\rm{charmless}}$, with the sensitivity differing among channels and partial waves. In Fig.~\ref{fig:directCP}, the magnitudes of the direct $CP$ asymmetries for the $N^*(1520)K_S$ and $N^*(1520)\bar K^{*0}$ modes increase with $\Lambda_{\rm charmless}$ and decrease with $\Lambda_{\rm charm}$. The $N^*(1535)$ modes show smaller absolute variations, although the asymmetry of the $N^*(1535)K_S$ mode can change sign. Among the partial-wave observables in Fig.~\ref{fig:partialCP}, $a_{CP}^{P}$ for $N^*(1520)K_S$ exhibits the largest absolute variation, becoming more negative as $\Lambda_{\rm charmless}$ increases and less negative as $\Lambda_{\rm charm}$ increases. Its $S$-wave counterpart follows the same trend with a smaller variation. For $N^*(1535)K_S$, both partial-wave asymmetries decrease as either parameter increases.

At the amplitude level, the dependence on the model parameters reflects the interference origin of direct and partial-wave $CP$ violation. These asymmetries depend on both the relative magnitudes and the relative weak and strong phases of the interfering amplitudes. In the representative channels considered here, the dominant charmed rescattering contribution to the tree amplitude is mainly controlled by $\Lambda_{\rm charm}$, whereas the charmless rescattering contribution to the penguin amplitude is controlled by $\Lambda_{\rm charmless}$. Variations of the two model parameters can therefore affect the relative magnitudes and strong phases of these contributions, rather than merely rescaling the full amplitude. Consequently, their effects need not cancel in the $CP$-asymmetry ratios, leading to the channel- and partial-wave-dependent behavior shown in Figs.~\ref{fig:directCP} and~\ref{fig:partialCP}.

\subsection{Asymmetry parameters}
For the $\Lambda_b\to N^*P/S$ and $\Lambda_b\to N^*V$
processes, the decay-asymmetry parameters characterize normalized combinations of helicity-amplitude magnitudes and interference terms.
These observables serve as sensitive probes of the angular distributions of the final-state particles, thereby providing deeper insight into the helicity structure of the weak Hamiltonian. Moreover, they offer valuable information on the angular correlations and dynamical features of the strong interaction. 
In the case of the $\Lambda_b \to N^*P/S$ processes, the decay asymmetry parameters are defined as \cite{Chen:2019hqi, Hong:2022prk}
\begin{equation}
    \begin{aligned}
        \alpha = \frac{\lvert H_{+1/2} \rvert^2 - \lvert H_{-1/2}\rvert^2}{\lvert H_{+1/2} \rvert^2 + \lvert H_{-1/2}\rvert^2}\ , \qquad \beta = \frac{2Im(H_{+1/2}H^*_{-1/2})}{\lvert H_{+1/2} \rvert^2 + \lvert H_{-1/2}\rvert^2}\ , \qquad \gamma = \frac{2Re(H_{+1/2}H^*_{-1/2})}{\lvert H_{+1/2} \rvert^2 + \lvert H_{-1/2}\rvert^2} \ ,
    \end{aligned}
\end{equation}
and the averaged asymmetry parameters and their corresponding $CP$ asymmetries are given by Ref. \cite{Donoghue:1986hh}
\begin{equation}
    \begin{aligned}
        \langle \alpha \rangle &= \frac{\alpha - \bar{\alpha}}{2}\ , \qquad 
        \langle \beta \rangle  = \frac{\beta - \bar{\beta}}{2}\ , \qquad
        \langle \gamma \rangle = \frac{\gamma + \bar{\gamma}}{2}\ , \\
        a_{CP}^{\alpha} &= \frac{\alpha + \bar{\alpha}}{2}\ , \qquad
        a_{CP}^{\beta}  = \frac{\beta + \bar{\beta}}{2}\ , \qquad
        a_{CP}^{\gamma} = \frac{\gamma - \bar{\gamma}}{2}\ .
    \end{aligned}
\end{equation}
\begin{table}[htbp]
\centering
\caption{Decay asymmetry parameters and their corresponding $CP$ asymmetries of $\Lambda_b \to N^*P/S$.}
\label{tab:asym_1535}

\renewcommand{\arraystretch}{1.2}
\setlength{\tabcolsep}{4pt} 

\resizebox{\textwidth}{!}{%
\begin{tabular}{lccccccccccc}
\hline\hline
\textbf{Decay channel} 
&  $\langle \alpha \rangle$ & $a_{CP}^{\alpha}(10^{-2})$ 
&  $\langle \beta \rangle$ & $a_{CP}^{\beta}(10^{-2})$ 
&  $\langle \gamma \rangle$ & $a_{CP}^{\gamma}(10^{-2})$ \\
\hline

$N^*(1535)K_S$ 

& $-0.10^{+0.01}_{-0.02}$
& $-0.07^{+0.42}_{-0.22}$

& $-0.47^{+0.00}_{-0.01}$
& $-0.25^{+0.07}_{-0.11}$

& $-0.87^{+0.00}_{-0.01}$
& $0.14^{+0.07}_{-0.07}$ \\ \hline

$ N^*(1535)K^*_0(700)$ 

& $-0.62^{+0.06}_{-0.02}$
& $19.30^{+7.47}_{-8.04}$

& $0.37^{+0.20}_{-0.20}$
& $39.93^{+7.53}_{-8.59}$

& $0.53^{+0.11}_{-0.16}$
& $5.67^{+3.12}_{-2.07}$ \\ \hline

$ N^*(1535)f_0(500)$ 
 
& $0.55^{+0.09}_{-0.13}$
& $-27.38^{+11.29}_{-15.92}$

& $0.52^{+0.06}_{-0.21}$
& $10.00^{+3.49}_{-14.81}$

& $0.57^{+0.11}_{-0.08}$
& $17.53^{+11.92}_{-7.60}$ \\ \hline

$ N^*(1535)f_0(980)$ 

& $-0.06^{+0.06}_{-0.17}$
& $40.0^{+1.98}_{-5.40}$

& $0.08^{+0.06}_{-0.06}$
& $35.39^{+3.32}_{-2.66}$

& $0.84^{+0.00}_{-0.01}$
& $-0.49^{+9.23}_{-4.40}$ \\ \hline

$ N^*(1520)K_S$ 

& $0.43^{+0.02}_{-0.01}$
& $0.99^{+0.68}_{-1.13}$

& $0.36^{+0.03}_{-0.05}$
& $0.19^{+0.28}_{-0.17}$

& $0.83^{+0.01}_{-0.01}$
& $0.44^{+0.52}_{-0.29}$ \\ \hline

$ N^*(1520)K^*_0(700)$ 

& $0.14^{+0.09}_{-0.13}$
& $4.45^{+3.55}_{-2.09}$

& $-0.11^{+0.05}_{-0.06}$
& $-2.45^{+2.03}_{-3.60}$

& $0.98^{+0.01}_{-0.02}$
& $-0.88^{+0.34}_{-0.24}$ \\ \hline

$ N^*(1520)f_0(500)$ 

& $-0.10^{+0.17}_{-0.14}$
& $36.47^{+0.00}_{-10.37}$

& $-0.85^{+0.04}_{-0.04}$
& $-7.15^{+6.45}_{-2.98}$

& $0.34^{+0.11}_{-0.10}$
& $-7.61^{+10.74}_{-9.60}$ \\ \hline

$ N^*(1520)f_0(980)$ 

& $-0.49^{+0.08}_{-0.04}$
& $10.27^{+4.41}_{-2.95}$

& $-0.77^{+0.04}_{-0.05}$
& $-5.61^{+2.45}_{-3.15}$

& $0.39^{+0.03}_{-0.02}$
& $1.69^{+4.93}_{-7.54}$ \\
\hline\hline
\end{tabular}
}
\end{table}
\begin{figure}[thb]
\begin{center}
\includegraphics[width=\columnwidth]{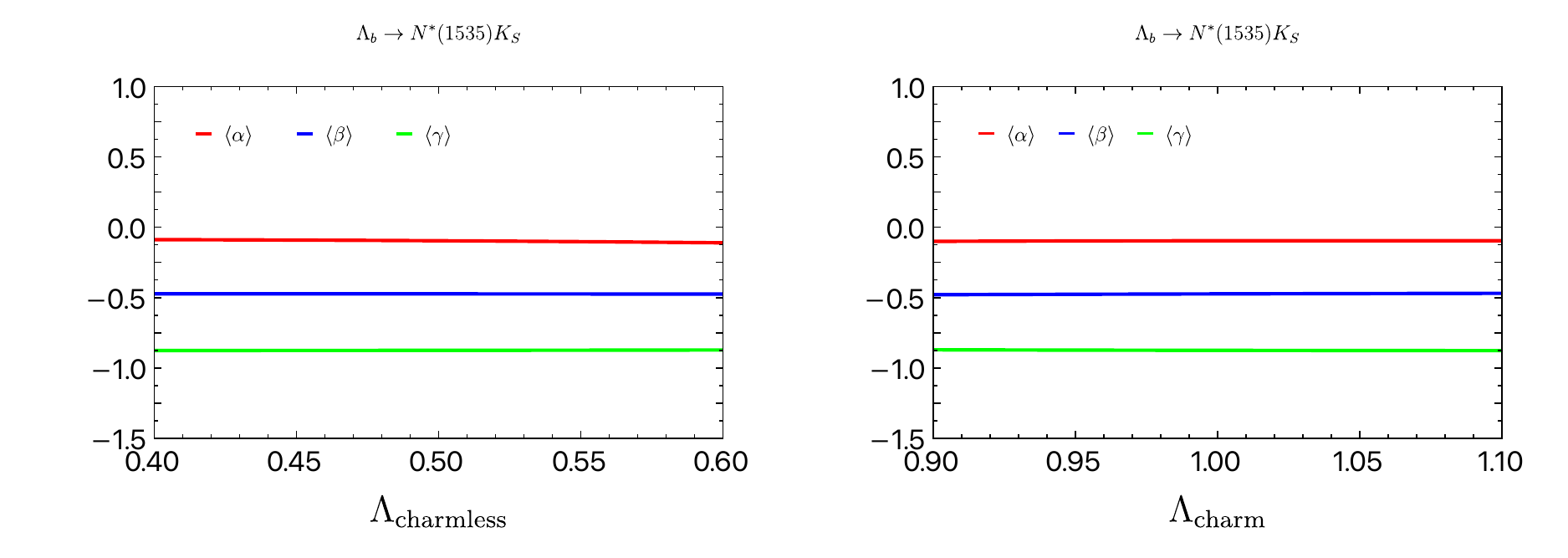}
\end{center}
\vspace{-20pt}
\caption{The dependence of the decay asymmetry parameters for $\Lambda_b \to N^*K_S$ on the model parameters $\Lambda_{\rm charmless}$ and $\Lambda_{\rm charm}$.}
\label{fig:apNKS}
\end{figure}
One can refer to Ref.~\cite{Qi:2025zna} for alternative definitions. The corresponding results are summarized in Table.~\ref{tab:asym_1535}.
The $CP$ asymmetries of the decay-asymmetry parameters are small for the $K_S$ modes, while several scalar-meson modes have much larger values, especially for $a_{CP}^{\alpha}$ and $a_{CP}^{\beta}$. In particular, the values of $a_{CP}^{\alpha}$ in the $N^*(1535)f_0(500)$, $N^*(1535)f_0(980)$, $N^*(1520)f_0(500)$, and $N^*(1520)f_0(980)$ channels reach the level of $\mathcal{O}(10^{-1})$. Figure ~\ref{fig:apNKS} shows that $\langle\alpha\rangle$, $\langle\beta\rangle$, and $\langle\gamma\rangle$ for $\Lambda_b\to N^*(1535)K_S$ depend only weakly on the two model parameters, $\Lambda_{\rm charmless}$ and $\Lambda_{\rm charm}$, within the considered parameter ranges. This weak dependence is consistent with a substantial
cancellation of the common normalization dependence in these ratios. The central value $\langle\gamma\rangle\simeq-0.87$ is consistent with the $P$-wave dominance of this channel shown in Table.~\ref{tab:Br_summary}.

\begin{figure}[thb]
\begin{center}
\includegraphics[width=\columnwidth]{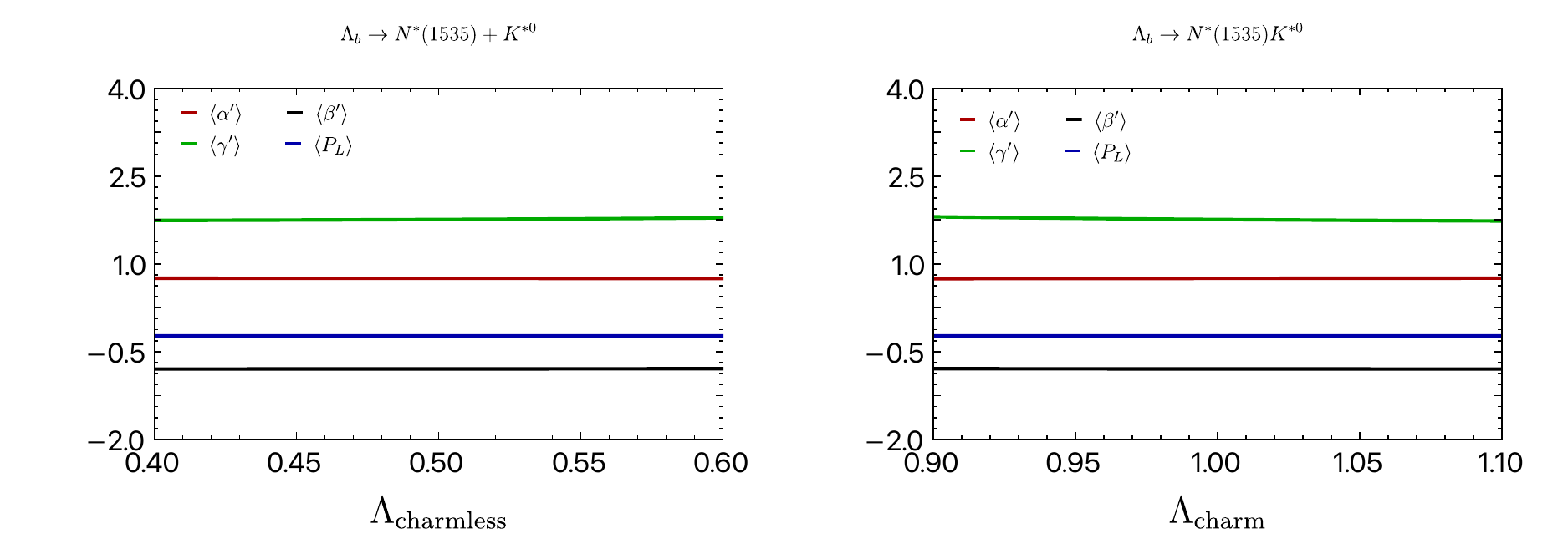}
\end{center}
\vspace{-20pt}
\caption{The dependence of the average of decay asymmetry parameters for $\Lambda_b \to N^*(1535)\bar{K}^{*0}$ on the model parameters $\Lambda_{\rm charmless}$ and $\Lambda_{\rm charm}$.}
\label{fig:apN1535V}
\end{figure}

In the case of the $\Lambda_b \to N^*(1535)V$ processes, the decay asymmetry parameters are defined as \cite{Chen:2019hqi, Hong:2022prk}
\begin{equation}
    \begin{aligned}
       & \alpha' = \frac{|H_{+1, +1/2}|^2 - |H_{-1, -1/2}|^2}{|H_{+1, +1/2}|^2 + |H_{-1, -1/2}|^2}, \quad 
        \beta' = \frac{|H_{0, +1/2}|^2 - |H_{0, -1/2}|^2}{|H_{0, +1/2}|^2 + |H_{0, -1/2}|^2}, \quad \\
       & \qquad \qquad \qquad \qquad \gamma' = \frac{|H_{+1, +1/2}|^2 + |H_{-1, -1/2}|^2}{|H_{0, +1/2}|^2 + |H_{0, -1/2}|^2} \ ,
    \end{aligned}
\end{equation}
and the longitudinal polarization of $N^*$ is characterized by
\begin{equation}
  P_L = \frac{|H_{+1, +1/2}|^2 - |H_{-1, -1/2}|^2 + |H_{0, +1/2}|^2 - |H_{0, -1/2}|^2}{|H_{+1, +1/2}|^2 + |H_{-1, -1/2}|^2 + |H_{0, +1/2}|^2 + |H_{0, -1/2}|^2} \ ,  
\end{equation}
these parameters are not independent and their relation is shown as follows
\begin{equation}
  P_L = \frac{\beta'+\alpha'\cdot \gamma'}{1+\gamma'} \ .
\end{equation}
The observable $\alpha'$ was first proposed by T. D. Lee and C. N. Yang in Ref.~\cite{Lee:1957qs}, which can be extracted from the angular distributions of the $N^*(1535)$ in the rest frame of $\Lambda_b$,
 \begin{equation}
     \begin{aligned}
         \frac{d\Gamma}{dcos\theta} \propto 1 + \mathcal{P}\alpha' cos\theta \ , 
     \end{aligned}
 \end{equation}
where $\mathcal{P}$ is the polarization of $\Lambda_b^0$ \cite{Wang:2022tcm}, and $\theta$ is the angle between the direction of the polarization of $\Lambda_b^0$  and the momentum of $N^*(1535)$. The other two parameters $\beta', \gamma'$ could be determined by measuring the polarization of $N^*(1535)$ in the final state \cite{Cronin:1963zb,Overseth:1967zz}. The decay asymmetry parameters for the channel $\Lambda_b \to N^*(1535)V$, calculated with model parameters $\Lambda_{\mathrm{charm}} = 1.0 \pm 0.1$ and $\Lambda_{\mathrm{charmless}} = 0.5 \pm 0.1$, are presented in Table. \ref{tab:apN1535V}. 
 For the $\bar K^{*0}$ mode, the central values $\langle\alpha'\rangle=0.75$ and $\langle\beta'\rangle=-0.794$ are sizable and have opposite signs. The $\rho(770)$ mode exhibits larger magnitudes of $a_{CP}^{\alpha'}$ and $a_{CP}^{\gamma'}$ than the $\bar K^{*0}$ mode.
Then we illustrate the dependence of the four decay asymmetry parameters for the process $\Lambda_b \to N^*(1535)\bar{K}^{*0}$ on the model parameters $\Lambda_{\mathrm{charm}}$ and $\Lambda_{\mathrm{charmless}}$ in Fig. \ref{fig:apN1535V}, and find that they are rather insensitive to the variations of the two model parameters.

\begin{table}[htbp]
\centering
\caption{Decay asymmetry parameters and their corresponding $CP$ asymmetries for $\Lambda_b \to N^*(1535)V$ processes.}
\label{tab:apN1535V}
\renewcommand{\arraystretch}{1.2}
\resizebox{0.7\textwidth}{!}{
\begin{tabular}{lcccccccccccccc}
\hline\hline
\textbf{Decay channel}  & $\langle \alpha^{\prime} \rangle $ & $a_{CP}^{\alpha^{\prime}}(10^{-2})$ & $\langle \beta^{\prime} \rangle$ & $a_{CP}^{\beta^{\prime}}(10^{-2})$  \\ 
\hline
$ N^*(1535)\bar{K}^{*0}$ 
& $0.75^{+0.01}_{-0.01}$
& $0.28^{+0.20}_{-0.24}$
 
& $-0.794^{+0.01}_{-0.01}$
& $0.36^{+0.27}_{-0.18}$
 \\
$ N^*(1535)\rho(770)$ 
& $0.60^{+0.07}_{-0.10}$
& $18.74^{+9.28}_{-8.71}$

& $-0.713^{+0.076}_{-0.083}$
& $1.35^{+2.06}_{-2.77}$ \\ \hline

\textbf{Decay channel} & $ \langle \gamma^{\prime} \rangle $ & $a_{CP}^{\gamma^{\prime}}(10^{-2})$ & $\langle P_L \rangle$ & $a_{CP}^{P_L}$ \\ \hline
$ N^*(1535)\bar{K}^{*0}$
 
& $1.76^{+0.05}_{-0.03}$
& $-0.13^{+0.03}_{-0.07}$

& $-0.23^{+0.01}_{-0.06}$
& $0.43^{+0.01}_{-0.01}$ \\ \hline
$ N^*(1535)\rho(770)$
 
& $0.96^{+0.37}_{-0.17}$
& $64.00^{+28.53}_{-19.46}$

& $0.35^{+0.09}_{-0.08}$
& $-0.13^{+0.04}_{-0.09}$ \\
\hline\hline
\end{tabular}
}
\end{table}

In the case of the $\Lambda_b \to N^*(1520) V$ processes, the decay width is defined as 
\begin{equation}
    \begin{aligned}
        \Gamma(\Lambda_b \to N^*(1520)V) &= \frac{p_c}{8\pi m_{\Lambda_b}^2} \frac{1}{2}( \lvert H_{+1, +3/2} \rvert^2 + \lvert H_{0, +1/2} \rvert^2 + \lvert H_{-1, -1/2}\rvert^2 + \lvert H_{+1, +1/2} \rvert^2 \\
        & + \lvert H_{0, -1/2} \rvert^2 + \lvert H_{-1, -3/2} \rvert^2) \ ,
    \end{aligned}
\end{equation}
and the two asymmetry parameters $A_T\sin(2\varphi_R)$ and $A_T\sin(\varphi_R)$ are introduced in Ref. \cite{Geng:2021lrc,Wang:2022fih,Wang:2024qff} as
\begin{equation}
    \begin{aligned}
        &A_T(\sin2\varphi_R) = Im(H_{+3/2, +1}H_{-1/2, -1}^*) - Im(H_{+1/2, +1}^*H_{-3/2, -1}) \ , \\
        &A_T(\sin\varphi_R) = Im(H_{+3/2, +1}H_{+1/2,0}^*)- Im(H_{-1/2, 0}^*H_{-3/2, -1}) \ .
    \end{aligned}
\end{equation}
As they correspond to the $T$-odd and parity-odd quantities, we define 
\begin{equation}
    \begin{aligned}
        &a_{CP}(\sin\varphi_R)  = \frac{A_T(\sin\varphi_R) + \bar{A}_T(\sin\varphi_R)}{\displaystyle\sum_{\lambda_f,\lambda_M} \lvert H_{\lambda_f,\lambda_M} \rvert^2 + \lvert \bar{H}_{\lambda_f,\lambda_M} \rvert^2} \ , \\
        &a_{CP}(\sin2\varphi_R)  = \frac{A_T(\sin2\varphi_R) + \bar{A}_T(\sin2\varphi_R)}{\displaystyle\sum_{\lambda_f,\lambda_M} \lvert H_{\lambda_f,\lambda_M} \rvert^2 + \lvert \bar{H}_{\lambda_f,\lambda_M} \rvert^2} \ .
    \end{aligned}
\end{equation}
Since the helicity amplitudes can be expressed as linear combinations of partial waves, the asymmetry parameters $A_T(\sin\varphi_R)$ and $A_T(\sin2\varphi_R)$ reduce to the imaginary part of the interference terms between partial waves with opposite parities. In addition to the $T$-odd observables $\sin\varphi_R$ and $\sin(2\varphi_R)$, the quantities $\cos\varphi_R$ and $\cos(2\varphi_R)$ are also valuable for constructing $T$-even $CP$-violating observables. Similar to $A_T\sin(\varphi_R)$ and $A_T\sin(2\varphi_R)$, we can define
\begin{equation}
    \begin{aligned}
        &A_T(\cos2\varphi_R) = Re(H_{+3/2, +1}H_{-1/2,-1}^*) + Re(H_{+1/2, +1}H_{-3/2,-1}^*) \ , \\
        &A_T(\cos\varphi_R) = Re(H_{+3/2, +1}H_{+1/2,0}^*) + Re(H_{-1/2,0}H_{-3/2, -1}^*) \ ,
    \end{aligned}
\end{equation}
and the other two $CP$-violating observables can be written as 
\begin{equation}
    \begin{aligned}
        &a_{CP}(\cos\varphi_R) = \frac{A_T(\cos\varphi_R) - \bar{A}_T(\cos\varphi_R)}{\displaystyle\sum_{\lambda_f,\lambda_M} | H_{\lambda_f,\lambda_M} |^2 + \lvert \bar{H}_{\lambda_f,\lambda_M} \rvert^2} \ , \\
        &a_{CP}(\cos2\varphi_R) = \frac{A_T(\cos2\varphi_R) - \bar{A}_T(\cos2\varphi_R)}{\displaystyle\sum_{\lambda_f,\lambda_M} | H_{\lambda_f,\lambda_M} |^2 + \lvert \bar{H}_{\lambda_f,\lambda_M} \rvert^2
        } \  ,
    \end{aligned}
\end{equation}
the two $CP$ asymmetries $a_{CP}(\cos\varphi_R)$ and $a_{CP}(\cos2\varphi_R)$ are complementary to $a_{CP}(\sin\varphi_R)$ and $a_{CP}(\sin2\varphi_R)$ \cite{Wang:2024qff}. The corresponding results are listed in Table. \ref{tab:asym_1520}. For the $\bar K^{*0}$ mode, their central magnitudes are of $\mathcal{O}(10^{-2})$ or smaller. For the $\rho(770)$ mode, $a_{CP}(\sin\varphi_R)$ has the largest central magnitude among the listed angular asymmetries.
\begin{table}[htbp]
\centering
\caption{$CP$ asymmetries of decay parameters in $\Lambda_b \to N^*(1520)V$ decays.}
\label{tab:asym_1520}
\renewcommand{\arraystretch}{1.2}
\begin{tabular}{lccccc}
\hline\hline
\textbf{Decay channel} & $a_{cp}(\sin\varphi_R)(\%)$ & $a_{cp}(\sin2\varphi_R)(\%)$ & $a_{cp}(\cos\varphi_R)(\%)$ & $a_{cp}(\cos2\varphi_R)(\%)$ \\ 
\hline
$ N^*(1520)\bar{K}^{*0}$ & $0.34^{+0.05}_{-0.08}$ & $-1.17^{+0.16}_{-0.24}$ & $0.17^{+0.05}_{-0.12}$ & $0.09^{+0.06}_{-0.03}$ \\
$ N^*(1520)\rho(770)$ & $-4.79^{+0.67}_{-0.50}$ & $0.11^{+0.34}_{-0.19}$ & $0.51^{+0.67}_{-0.10}$ & $-0.95^{+0.24}_{-1.40}$ \\
\hline\hline
\end{tabular}
\end{table}

\section{CONCLUSION}
In this work, we studied the charmless nonleptonic two-body decays $\Lambda_b\to N^*M$ within the final-state rescattering framework, where the final state consists of an excited nucleon $N^*$ and a meson \(M\), with $M =K_S, K^*_0(700)$, $f_0(500,980)$, $\rho(770)$, $\bar{K}^{*0}$, and $\phi$ . We adopted the two model parameters, $\Lambda_{\rm charm}$ and $\Lambda_{\rm charmless}$, with their values taken from Ref.~\cite{Duan:2024zjv}, and evaluated the relevant hadronic triangle diagrams with both their absorptive and dispersive parts retained.
We then predicted the branching ratios, direct and partial-wave $CP$ asymmetries, and decay-asymmetry parameters, together with complementary $T$-odd and $T$-even angular $CP$ asymmetries for the $N^*(1520)V$ modes. The $\bar K^{*0}$ and $\rho^0$ modes generally have the largest branching ratios, followed by the $K_S$ modes, while the scalar-meson modes and especially the $\phi$ modes are suppressed. The partial-wave results show that $\Lambda_b^0\to N^*(1535)K_S$ is dominated by
the $P$-wave contribution and illustrate how
partial-wave rate fractions determine the weights of the corresponding asymmetries in the total direct $CP$ asymmetry.
By studying these processes, we can also provide theoretical input for analyses of the $N^*M$ quasi-two-body components in experimentally accessible multibody $\Lambda_b$ decays. These results may be useful for investigating resonance-dominated regions and the corresponding $CP$ asymmetries in $\Lambda_b$ four-body decay channels, such as $\Lambda_b^0\to p\pi^-\pi^+\pi^-$, in future LHCb measurements.

The dependence on the two model parameters has also been examined. Within the considered ranges of the two model parameters, 
the branching ratios depend strongly on $\Lambda_{\rm charm}$ and much less strongly on $\Lambda_{\rm charmless}$. The direct and partial-wave $CP$ asymmetries can exhibit a visible dependence on both parameters, whereas several averaged decay-asymmetry parameters vary only weakly over the considered ranges. These physical observables provide complementary information on the normalization, interference, and relative helicity structure of the decay amplitudes.

Further reducing the theoretical uncertainties in the predicted physical observables is an important direction for future work.
 In addition to the uncertainties associated with the two model parameters, \(\Lambda_{\rm charm}\) and \(\Lambda_{\rm charmless}\) ,  additional uncertainties arise from the heavy-to-light transition form factors, the strong coupling constants involving excited nucleons, the use of approximate flavor symmetries for some presently unknown charmed-hadron couplings, and the omission of rescattering contributions involving unavailable scalar-meson vertices.
More precise determinations of these hadronic inputs will therefore be important for improving the precision of the theoretical predictions.

\paragraph{ACKNOWLEDGEMENTS}
 The authors are very grateful to Jian-Peng Wang, Cai-Ping Jia, Zhu-Ding Duan for helpful discussions.  This work is supported by the National Natural Science Foundation of China under Grant No. 12335003, No. 12522506 and No. 12375086, and by the Fundamental Research Funds for the Central Universities under No. lzujbky-2023-stlt01 and No.lzujbky-2024-oy02.

\newpage
\appendix

\section{EFFECTIVE LAGRANGIAN}\label{EL}
\begin{itemize}
    \item The effective Lagrangians for pseudoscalar meson octet $P_8$, vector meson $V$, charmed baryon sextets $B_6$ and antitriplets $B_{\bar{3}}$:
\begin{equation}
    \begin{aligned}
        &\mathcal{L}_{VPP} = \frac{i g_{VPP}}{\sqrt{2}}Tr[V^{\mu}[P, \partial_{\mu} P]]\ ,\\
	&\mathcal{L}_{VVV} = \frac{i g_{VVV}}{\sqrt{2}}Tr[(\partial_{\nu} V_{\mu}V^{\mu} - V^{\mu}\partial_{\nu} V_{\mu})V^{\nu}] \ , \\
	&\mathcal{L}_{PB_{6}B_{6}} = g_{PB_{6}B_{6}}Tr[\bar{B}_{6}i\gamma_{5}PB_{6}],\\
	&\mathcal{L}_{VB_{6}B_{6}} = f_{1VB_{6}B_{6}}Tr[\bar{B}\gamma_{\mu}V^{\mu}B] + \frac{f_{2VB_{6}B_{6}}}{m_{6} + m_{6}}Tr[\bar{B}\sigma_{\mu \nu}\partial^{\mu}V^{\nu}B],\label{A.4}\\
	&\mathcal{L}_{PB_{6}B_{\bar{3}}} = g_{PB_{6}B_{\bar{3}}}Tr[\bar{B}_{6}i\gamma_{5}PB_{\bar{3}}] + H.c. \ ,  \\
	&\mathcal{L}_{VB_{6}B_{\bar{3}}} = f_{1VB_{6}B_{\bar{3}}}Tr[\bar{B}_{6}\gamma_{\mu}V^{\mu}B_{\bar{3}}] + \frac{f_{2VB_{6}B_{\bar{3}}}}{m_{3} + m_{6}}Tr[\bar{B}_{6}\sigma_{\mu \nu}\partial^{\mu} V^{\nu} B_{\bar{3}}] + H.c. \ .
    \end{aligned}
\end{equation}

\item The effective Lagrangians involving $D/D^*$ mesons:
\begin{equation}
	\begin{aligned}
		&\mathcal{L}_{D^*DP} = -ig_{D^*DP}(D^i\partial^{\mu}P_{ij}D_{\mu}^{*j+} - D_{\mu}^{*i}\partial^{\mu}P_{ij}D^{j+}) \ , \\
		&\mathcal{L}_{D^*D^*P} = \frac{1}{2}g_{D^*D^*P}\epsilon_{\mu \nu \alpha \beta}D^{*\mu}_i\partial^{\nu}P^{ij} \overleftrightarrow{\partial}^{\alpha}D_j^{*\beta+} \ , \\
		&\mathcal{L}_{DDV} = -ig_{DDV}D_i\overleftrightarrow{\partial}_{\mu}D^{j+}(V^{\mu})^i_j \ , \\
		&\mathcal{L}_{D^*DV} = -2f_{D^*DV}\epsilon_{\mu \nu \alpha \beta}(\partial^{\mu}V^{\nu})^i_j(D_i\overleftrightarrow{\partial}^{\alpha}D^{*\beta j+} - D_i^{*\beta}\overleftrightarrow{\partial}^{\alpha}D^{j+}) \ ,  \\
		&\mathcal{L}_{D^*D^*V} = ig_{D^*D^*V}D^{*\nu +}_i\overleftrightarrow{\partial}_{\mu}D_{\nu}^{*j}(V^{\mu})^i_j + 4if_{D^*D^*V}D^{*+}_{i\mu}(\partial^{\mu}V^{\nu} - \partial^{\nu}\partial^{\mu})^i_jD_{\nu}^{*j} \ .
	\end{aligned}
\end{equation}

\item The effective Lagrangians involving $N^*(1535,1520)$:
\begin{equation}\label{eq:LN*}
    \begin{aligned}
        \mathcal{L}_{\pi NN^*}^{1520} & = \frac{f_{\pi NN^*}^{1520}}{m_{\pi}}\bar{\psi}\chi^{\dagger}\gamma_5\partial_{\mu}\vec{\phi} \cdot \vec{\tau}\chi \psi_{\mu} + H.c. \ , \\
        \mathcal{L}_{\pi NN^*}^{1535} &= \frac{if_{\pi NN^*}^{1535}}{m_{\pi}}\bar{\psi}\chi^{\dagger}\gamma_{\mu}\partial_{\mu}\vec{\phi}\cdot \vec{\tau}\chi\psi_{N^*} + h.c. \ , \\
        \mathcal{L}_{\omega NN^*}^{1520} &= \frac{ig_{\omega NN^*}^{1520}}{m_{\omega}^2}\bar{\psi}_N\sigma_{\mu \nu}\partial_{\nu}\partial_{\kappa}\omega_{\mu}\psi_{\kappa} + H.c. \ , \\
        \mathcal{L}_{\omega NN^*}^{1535} &= -\frac{ig_{\omega NN^*}^{1535}}{m_{\omega}^2}\bar{\psi}_N\gamma_5[\gamma_{\mu}\partial^2 - (m^* + m_N)\partial_{\mu}]\omega_{\mu}\psi_{N^*} + H.c. \ , \\
        \mathcal{L}_{\rho NN^*}^{1520} &= \frac{ig_{\rho NN^*}^{1520}}{m_{\rho}^2}\bar{\psi}\chi^{\dagger}\sigma_{\mu \nu}\partial_{\nu}\partial_{\kappa}\vec{\tau}\cdot \vec{\rho}_{\mu}\chi\psi_{\kappa} + H.c. \ , \\
        \mathcal{L}_{\rho NN^*}^{1535} &= -\frac{ig_{\rho NN^*}^{1535}}{m_{\rho}^2}\bar{\psi}\chi^{\dagger}\gamma_5[\gamma_{\mu}\partial^2 - (m^* + m_N)\partial_{\mu}]\vec{\tau}\cdot \vec{\rho}_{\mu}\chi \psi_{N^*} + H.c. \ .
    \end{aligned}
\end{equation}
\end{itemize}

\section{FEYNMAN RULES FOR STRONG INTERACTION VERTICES}\label{FR}
The corresponding Feynman rules for strong interaction vertices used in this work are as follows:
\begin{equation}
    \begin{aligned}
 		&\langle P(p_3)D(k)|i\mathcal{L}_{D^*DP}|D^{*}(p_1)\rangle = i^3g_{D^*DP}\epsilon_{\mu}(p_1)p_3^{\mu} \ , \\
         &\langle P(p_3)D^{*}(k)|i\mathcal{L}_{D^*D^*P}|D^{*}(p_1) \rangle 
 		= i^3 \frac{1}{2}g_{D^*D^*P}\epsilon_{\mu \nu \alpha \beta}\epsilon^{\mu}(p_1)[p_3^{\nu}p_3^{\alpha}\epsilon^{*\beta}(k) - p_3^{\nu}k^{\alpha}\epsilon^{*\beta}(k)] \ , \\
         &\langle P(p_3)|i\mathcal{L}_{VPP}|P(k)V(p_1)\rangle
 		= i^3\frac{g_{VPP}}{\sqrt{2}}\epsilon^{*\mu}(p_1)(p_3 + k)_{\mu}\ , \\
         &\langle P(p_3)|i\mathcal{L}_{VVP}|V(p_1)V(k) \rangle
 		= i^3\frac{4g_{VVP}}{f_P}\epsilon^{\mu \nu \alpha \beta}p_{1\mu}\epsilon_{\nu}(p_1)k_{\alpha}\epsilon_{\beta}(k) \ , \\
 		&\langle K^*_0(700)(p_3)D(k)|i\mathcal{L}_{K^*_0(700)DD}|D_s(p_1)\rangle = -i^3\gamma_{\kappa K \pi}p_{1\mu}k^{\mu},\\ 
   		& \langle \sigma(p_3)|i\mathcal{L}_{\sigma DD}|D(p_1)D(k)\rangle = -i^3 \frac{\gamma_{\sigma DD}}{\sqrt{2}}p_1 \cdot k,\\
  		&\langle \sigma(p_3)\Lambda_c(k)|i\mathcal{L}_{\sigma NN}|\Lambda_c(p_2)\rangle = ig_{\sigma NN}\bar{u}(k,s_k)u(p_2,s_2) \ , \\
 		& \langle f_0(980)(p_3)|i\mathcal{L}_{f_0DD}|D(p_1)D(k)\rangle = i\frac{1}{2}g_{f_0 \pi \pi} = i\frac{1}{2}g_{f_0 DD} \ , \\
        &\langle N^*(1520)(p_4)|i\mathcal{L}_{\Lambda_c DN^*}|\Lambda_c(p_2)D(k) \rangle
 		= -i^2\frac{f_{\pi NN*}^{1520}}{m_{\pi}}\bar{u}_{\mu}(p_4, s_4)\gamma_5k^{\mu}u(p_2, s_2)\ ,  \\
         &\langle N^*(1535)(p_4)|i\mathcal{L}_{\Lambda_c DN^*}|\Lambda_c(p_2)D(k)\rangle 
 		=  -i^3 \frac{f_{\pi NN^*}^{1535}}{m_{\pi}}\bar{u}(p_4, s_4)\gamma_{\mu}k^{\mu}u(p_2,s_2)\ ,  \\
        &\langle N^*(1520,p_4)|i\mathcal{L}_{V NN^*}^{1520}|\Lambda_c(p_2)D^{*}(k)\rangle 
 		= i^4\frac{g_{\rho NN^*}^{1520}}{m_{\rho}^2}\bar{u}_{\alpha}(p_4, s_4)\sigma_{\mu \nu}k^{\nu}k^{\alpha}\epsilon^{*\mu}(k)u(p_2, s_2)\ ,  \\
        &\langle N^*(1535)(p_4)|i\mathcal{L}_{\rho NN^*}^{1535}|\Lambda_c(p_2)D^{*}(k)\rangle
 		=-i^2 \frac{g_{\rho NN^*}^{1535}}{m_{\rho}^2}\bar{u}(p_4, s_4)[\gamma_{\mu}(-k^2) - (m_2 + m_4) \\ 
         & \qquad \qquad \qquad \qquad \qquad \qquad \qquad \qquad \times (-ik_{\mu})] \times \gamma_5\epsilon^{\mu}(k)u(p_2, s_2)\ ,  \\
         &\langle N^{*}(1520)(p_4)V(k)|i\mathcal{L}_{V NN^*}^{1520}|B_8(p_2)\rangle
 		= -i^4 \frac{g_{\rho NN^*}^{1520}}{m_{\rho}^2}\bar{u}_{\kappa}(p_4,s_4)\sigma_{\mu \nu}k^{\nu}k^{\kappa}\epsilon^{*\mu}(k)u(p_2,s_2)\ ,  \\
 		&\langle N^{*}(1535)(p_4)V(k)|i\mathcal{L}_{V NN^*}^{1535}|B_8(p_2)\rangle  = i^2 \frac{g_{\rho NN^*}^{1535}}{m_{\rho}^2}\bar{u}(p_4,s_4)[\gamma_{\mu}(-k^2) - i(m_2 + m_4)k_{\mu}] \\ 
         & \qquad \qquad \qquad\qquad \qquad \qquad\qquad \qquad  \times \gamma_5\epsilon^{*\mu}(k)u(p_2,s_2)\ ,\\
         &\langle V(p_3)D(k)|i\mathcal{L}_{DDV}|D_s(p_1)\rangle  
 		= -i^3g_{DDV}(k_{\mu}+ p_{1\mu})\epsilon^{*\mu}(p_3) \ , \\
        &\langle V(p_3)D(k)|i\mathcal{L}_{D^*DV}|D_s^{*-}(p_1)\rangle 
 		= -2i^3f_{D^*DV}\epsilon_{\mu \nu \alpha \beta}(p_1)p_3^{\mu}(k + p_1)^{\alpha}\epsilon^{*\nu}(p_3)\epsilon^{\beta}(p_1) \ , \\
        & \langle V(p_3)D^{*}(k)|i\mathcal{L}_{D^*D^*V}|D_s^{*}(p_1)\rangle  
 		= i^3g_{D^*D^*V}(k + p_1)_{\mu}\epsilon^{*\nu}(k)\epsilon_{\nu}(p_1)\epsilon^{*\mu}(p_3) \\
         & \qquad \qquad \qquad\qquad \qquad \qquad \qquad + 4i^3 f_{D^*D^*V}\epsilon^*_{\mu}(k)[p_3^{\mu}\epsilon^{*\nu}(p_3) - p_3^{\nu}\epsilon^{*\mu}(p_3)]\epsilon_{\nu}(p_1) \ , \\
         &\langle V(p_3)|i\mathcal{L}_{VVV}|V(p_1)V(k)>
 		= i^3\frac{g_{VVV}}{\sqrt{2}}[-k_{\nu}\epsilon_{\mu}(k)\epsilon^{*\mu}(p_3)\epsilon^{\nu}(p_1) -\epsilon_{\mu}(k)p_{3\nu}\epsilon^{*\mu}(p_3) \\
 		& \qquad \qquad \qquad\qquad \qquad \qquad \qquad \times  \epsilon^{\nu}(p_1) + p_{3\nu}\epsilon^*_{\mu}(p_3)\epsilon^{\mu}(p_1)\epsilon^{\nu}(k) + \epsilon^{*\mu}(p_3)p_{1\nu} \\
 		& \qquad \qquad \qquad\qquad \qquad \qquad \qquad \times \epsilon_{\mu}(p_1)\epsilon^{\nu}(k)] \ , \\
         & \langle N^*(1535)(p_4)|i\mathcal{L}_{\pi NN^*}^{1535}|\Lambda_c(p_2)D(k)\rangle = i^3 \frac{f_{\pi NN^*}^{1535}}{m_{\pi}}\bar{u}(p_4,s_4)\gamma_{\mu}k^{\mu}u(p_2,s_2)\ ,  \\
 	\end{aligned}
 \end{equation}

\begin{equation}
    \begin{aligned}
        & \langle N^*(1520)(p_4)|i\mathcal{L}_{\pi NN^*}^{1520}|\Lambda_c(p_2)D(k)\rangle = i^2 \frac{f_{\pi NN^*}^{1520}}{m_{\pi}}\epsilon^*_{\mu}(p_4)\bar{u}(p_4,s_4)\gamma_{5}k^{\mu}u(p_2,s_2) \ , \\
        &\langle N^*(1535)(p_4)|i\mathcal{L}_{\rho NN^*}^{1535}|\Lambda_c(p_2)D^{*}(k)\rangle 
 		= -i\frac{g_{\rho NN^*}^{1535}}{m_{\rho}^2}\bar{u}(p_4,s_4)[\gamma_{\mu}k^2 - i(m_2 + m_4)k_{\mu}] \\
 		& \qquad \qquad \qquad\qquad \qquad \qquad \qquad ~~~~~\times \gamma_5\epsilon^{\mu}(k) u(p_2,s_2), \\
 		&\langle N^*(1520)(p_4)|i\mathcal{L}_{\rho NN^*}^{1520}|\Lambda_c(p_2)D^{*}(k)\rangle
 		= i^3 \frac{g_{\rho NN^*}^{1520}}{m_{\rho}^2}\epsilon_{\kappa}^*(p_4)\bar{u}(p_4,s_4)\sigma_{\mu \nu}k^{\nu}k^{\kappa}\epsilon^{\mu}(k)\\ 
         & \qquad \qquad \qquad\qquad \qquad \qquad\qquad \qquad  \times u(p_2,s_2) \ , \\
        &\langle N^*(1520)(p_4)|i\mathcal{L}_{V NN^*}^{1520}|\Lambda_c(p_2)D^{*}(k)\rangle 
 		= i^4\frac{g_{\rho NN^*}^{1520}}{m_{\rho}^2}\bar{u}_{\alpha}(p_4, s_4)\sigma_{\mu \nu}k^{\nu}k^{\alpha}\epsilon^{*\mu}(k)u(p_2, s_2)\ ,  \\
        &\langle N^*(1535)(p_4)|i\mathcal{L}_{\rho NN^*}^{1535}|\Lambda_c(p_2)D^{*}(k)\rangle
 		=-i^2 \frac{g_{\rho NN^*}^{1535}}{m_{\rho}^2}\bar{u}(p_4, s_4)[\gamma_{\mu}(-k^2) - (m_2 + m_4)\\ 
         & \qquad \qquad \qquad\qquad \qquad \qquad\qquad \qquad  \times (-ik_{\mu})] \gamma_5\epsilon^{\mu}(k)u(p_2, s_2)\ ,  \\
        &\langle N^{*}(1520)(p_4)V(k)|i\mathcal{L}_{V NN^*}^{1520}|B_8(p_2)\rangle
 		= -i^4 \frac{g_{\rho NN^*}^{1520}}{m_{\rho}^2}\bar{u}_{\kappa}(p_4,s_4)\sigma_{\mu \nu}k^{\nu}k^{\kappa}\epsilon^{*\mu}(k)u(p_2,s_2)\ ,  \\
 		&\langle N^{*}(1535)(p_4)V(k)|i\mathcal{L}_{V NN^*}^{1535}|B_8(p_2)\rangle  = i^2 \frac{g_{\rho NN^*}^{1535}}{m_{\rho}^2}\bar{u}(p_4,s_4)[\gamma_{\mu}(-k^2) - i(m_2 + m_4)k_{\mu}] \\ 
        & \qquad \qquad \qquad\qquad \qquad \qquad \qquad \qquad \times \gamma_5\epsilon^{*\mu}(k)u(p_2,s_2)\ .
    \end{aligned}
\end{equation}

\section{FULL EXPRESSIONS OF AMPLITUDES}\label{FA}
We present the expressions for the full amplitudes of $\Lambda_b$ decay channels

\begin{itemize}
    \item $\Lambda_b \to N^*_{1520,1535}K_S$:
    \begin{equation}
        \begin{aligned}
            \mathcal{A}(\Lambda_b &\to N^*_{1520,1535} K_S) = \mathcal{S}(\Lambda_b \to N^*K_S) + \mathscr{M}[D_s^{*-}, \Lambda_c^+; D^-] + \mathscr{M}[D_s^{-}, \Lambda_c^+; D^{*-}] \\
            & + \mathscr{M}[D_s^{*-}, \Lambda_c^+; D^{*-}] + \mathscr{M}[K^{*-}, p; \pi^+] + \mathscr{M}[K^-,p ;\rho^+] + \mathscr{M}[K^{*-},p ;\rho^+] \\
            & + \mathscr{M}[\bar{K}^{*0},n ;\pi^0] + \mathscr{M}[\bar{K}^{*0},n ;\rho^0]  + \mathscr{M}[\bar{K}^{*0}, n; \omega] + \mathscr{M}[\bar{K}^0,n; \rho^0] \\
            & + \mathscr{M}[\bar{K}^0,n ; \omega] + \mathscr{M}[\rho^0, \Lambda^0 ; \bar{K}^0] + \mathscr{M}[\omega, \Lambda^0 ;\bar{K}^0] + \mathscr{M}[\phi, \Lambda^0;\bar{K}^{*0}] \\
            & + \mathscr{M}[\pi^0, \Lambda^0; \bar{K}^{*0}] + \mathscr{M}[\rho^0, \Lambda^0 ; n] + \mathscr{M}[\omega, \Lambda^0; n] + \mathscr{M}[\pi^0, \Lambda^0; n] \ . 
        \end{aligned}
    \end{equation}
    \item $\Lambda_b \to N^*_{1520,1535}\bar{K}^{*0}$:
    \begin{equation}
        \begin{aligned}
            \mathscr{A}(\Lambda_b & \to N^*_{1520,1535}\bar{K}^{*0}) = \mathcal{S}(\Lambda_b \to N^*\bar{K}^{*0}) + \mathscr{M}[D_s^-,\Lambda_c^+; D^-] + \mathscr{M}[D_s^{*-},\Lambda_c^+; D^-] \\& 
            + \mathscr{M}[D_s^-,\Lambda_c^+; D^{*-}] + \mathscr{M}[D_s^{*-},\Lambda_c^+; D^{*-}]
            + \mathscr{M}[K^-,p; \pi^+] + \mathscr{M}[K^-,p; \rho^+] \\
            & + \mathscr{M}[K^{*-},p;\rho^+] + \mathscr{M}[K^{*-},p; \pi^+]  + \mathscr{M}[\bar{K}^0,n;\pi^0] + \mathscr{M}[\bar{K}^0,n; \rho^0] \\
            &+ \mathscr{M}[\bar{K}^0, n; \omega]  + \mathscr{M}[\bar{K}^{*0},n; \pi^0]  + \mathscr{M}[\bar{K}^{*0},n; \rho^0]  + \mathscr{M}[\bar{K}^{*0},n;\omega] \\
            & + \mathscr{M}[\pi^0,\Lambda^0;\bar{K}^0] + \mathscr{M}[\rho^0, \Lambda^0;\bar{K}^0] + \mathscr{M}[\omega,\Lambda^0; \bar{K}^0] + \mathscr{M}[\phi, \Lambda^0; \bar{K}^0] \\
            &+ \mathscr{M}[\pi^0, \Lambda^0; \bar{K}^{*0}] + \mathscr{M}[\rho^0,\Lambda^0; \bar{K}^{*0}] + \mathscr{M}[\omega, \Lambda^0; \bar{K}^{*0}] + \mathscr{M}[\phi, \Lambda^0; \bar{K}^{*0}] \ .  
        \end{aligned}
    \end{equation}
\end{itemize}

\begin{itemize}
    \item $\Lambda_b \to N^*_{1520,1535}K^*_0(700)$: 
    \begin{equation}
    	\begin{aligned}
    		& \mA(\Lambda_b \to N^*_{1520,1535}K^*_0(700)) = \mathcal{S}(\Lambda_b \to N^*K^*_0(700)) + \mM[D_s^-, \Lambda_c^+; D^-] \\
    		& \qquad \qquad + \mM[K^-,p; \pi^+] + \mM[\bar{K}^0, n; \pi^0]  + \mM[\pi^0, \Lambda^0; \bar{K}^0] + \mM[\pi^0, \Lambda^0;n] \\
    		& \qquad \qquad+ \mM[\rho^0, \Lambda^0; n] + \mM[\omega, \Lambda^0; n] \ .
    	\end{aligned}
    \end{equation}
    \item $\Lambda_b \to N^*_{1520,1535} f_0(500,980)$ : 
    \begin{equation}
    	\begin{aligned}
    		&\mA(\Lambda_b \to N^*_{1520,1535}f_0(500,980)) = \mathcal{S}(\Lambda_b \to N^*f_0(500,980)) + \mM[D^-, \Lambda_c^+; D^+] \\
    		& \qquad \qquad \qquad + \mM[D^-, \Lambda_c^+; \Lambda_c^+] + \mM[D^{*-}, \Lambda_c^+; \Lambda_c^+] + \mM[\pi^-, p; \pi^+] + \mM[\pi^-,p; p] \\
    		& \qquad \qquad \qquad + \mM[\rho^-, p; p] + \mM[\pi^0,n; \pi^0] + \mM[\pi^0, n; n] + \mM[\rho^0,n ; n] + \mM[\omega, n; n] \\
    		& \qquad \qquad \qquad + \mM[K^0, \Lambda^0; \bar{K}^0] + \mM[K^0, \Lambda^0; \Lambda^0] + \mM[K^{*0}, \Lambda^0; \Lambda^0] \ .
    	\end{aligned}
    \end{equation} 
\end{itemize}

\begin{itemize}
	\item $\Lambda_b \to N^*_{1520,1535}\rho^0(770): $
	\begin{equation}
		\begin{aligned}
			& \mA(\Lambda_b \to N^*_{1520,1535}\rho^0(770)) = \mathcal{S}(\Lambda_b \to N^*\rho^0(770)) + \mM[D^-,\Lambda_c^+; D^{-}] \\
			& \qquad \qquad + \mM[D^-, \Lambda_c^+; D^{*-}]   + \mM[D^{*-}, \Lambda_c^+; D^-] + \mM[D^{*-}, \Lambda_c^+; D^{*-}] \\
			& \qquad \qquad+ \mM[D^-, \Lambda_c^+; \Sigma^+_c] + \mM[D^{*-}, \Lambda_c^+; \Sigma^+_c] + \mM[\pi^-, p; \pi^+] \\
			& \qquad \qquad + \mM[\rho^-, p; \rho^+] + \mM[\pi^-, p; p] + \mM[\rho^-, p; p] + \mM[\omega, n; \pi^0] \\
			& \qquad \qquad + \mM[\pi^0,n; \omega] + \mM[\pi^0, n; n] + \mM[\rho^0, n; n] + \mM[\omega, n; n] \\
			& \qquad \qquad + \mM[K^0, \Lambda^0; \bar{K}^0] + \mM[K^0, \Lambda^0; \bar{K}^{*0}] + \mM[K^{*0}, \Lambda^0; \bar{K}^0] \\
			& \qquad \qquad + \mM[K^{*0},\Lambda^0; \bar{K}^{*0}] \ .
		\end{aligned}
	\end{equation}
	\item $\Lambda_b \to N^*_{1520,1535}\phi$ : 
	\begin{equation}
		\begin{aligned}
			& \mA(\Lambda_b \to N^*_{1520,1535}\phi) =\mathcal{S}(\Lambda_b \to N^*\phi) + \mM[K^0, \Lambda^0; \bar{K}^0] + \mM[K^0,\Lambda^0; \bar{K}^{*0}] \\
			& \qquad ~ + \mM[K^{*0},\Lambda^0; \bar{K}^0] + \mM[K^{*0},\Lambda^0; \bar{K}^{*0}]  + \mM[K^0, \Lambda^0; \Lambda^0] + \mM[K^{*0}, \Lambda^0; \Lambda^0],
		\end{aligned}
	\end{equation}
\end{itemize}
where $\mathcal{S}$ denotes the short-distance contributions.

\end{document}